\documentclass[aps,pra,twocolumn,amsmath,amssymb]{revtex4-2}

\usepackage{amsmath}
\usepackage{graphicx}
\usepackage{dcolumn}
\usepackage{bm}
\usepackage{xcolor}
\usepackage[normalem]{ulem}                 
\usepackage{booktabs}
\usepackage{hyperref}

\usepackage{subfigure}

\begin{document}

\title{Dynamics of a Single-Domain Dissipative Magnetic Gyroscope in Electric and Magnetic Fields}

\author{Y. B. Band$^{1,2,3}$, Igor Kuzmenko$^{1,2}$}

\affiliation{
  $^1$Department of Physics,
  Ben-Gurion University of the Negev,
  Beer-Sheva 84105, Israel
  \\
  $^2$Department of Chemistry,
  Ben-Gurion University of the Negev,
  Beer-Sheva 84105, Israel
  \\
  $^3$The Ilse Katz Center for Nano-Science,
  Ben-Gurion University of the Negev,
  Beer-Sheva 84105, Israel
  }

\date{\today}

\begin{abstract}
We formulate the coupled spin--rotation dynamics of a single-domain symmetric top ferromagnetic needle (a magnetic gyroscope) in static magnetic and electric fields. The conservative Hamiltonian contains the rotational kinetic energy, the magnetocrystalline and shape anisotropy, the Zeeman energy, the electrostatic energy of the dipole induced in the conducting needle, the linear magnetoelectric coupling, and the relativistic motional (spin-orbit) coupling. Gilbert damping is introduced, and in an electric field, Debye relaxation of the induced electric dipole of the conducting needle also results in dissipation.  We give the equations of motion in the laboratory and body-fixed frames, with singularity-free quaternion kinematics. For a 120-nm-long single-domain cobalt needle in a 0.5~T field we find: (i) because the anisotropy field ($\approx 1.3$~T) exceeds the applied magnetic field, the spin reverses direction several times during the initial Gilbert-damping transient, and the parity of the number of reversals determines whether the easy axis ends up parallel or antiparallel to the spin; (ii) after the transient the needle librates about the field---precession accompanied by nutation at twice the precession frequency---with a period of about 20~ns, consistent with the libration period $2\pi(I_1/\mu B)^{1/2}$, while the Zeeman and rotational kinetic energies oscillate out of phase and the total energy is conserved; (iii) a static electric field $\mathbf E$ exerts on the conducting needle a shape-dependent polarization torque that tends to align the long axis of the needle with $\mathbf E$.
\end{abstract}

\maketitle

\section{Introduction}

The mechanical manipulation of macroscopic spin systems offers rich opportunities for studying quantum macroscopicity, precision magnetometry, and mechanical torque detection \cite{Kimball2016, Rusconi2017, Band2018, Wang2019, Fadeev2021, Fadeev2021b, Vinante2021, Kalia2024, Kimball2026, Rodriguez2025}. A single-domain ferromagnet possesses a macro-spin  of magnitude $S = N_s\hbar/2$, where $N_s$ is the net number of uncompensated electron spins coupled to its crystalline lattice or geometric easy axis $\hat{\mathbf{n}}$ via magnetocrystalline and shape anisotropy \cite{Kimball2016, Rusconi2017, Band2018, Wang2019, Fadeev2021, Fadeev2021b, Kimball2026, Chudnovsky1998}.  We treat the magnetic needle as a rigid symmetric top with principal moments of inertia $I_1, I_2 = I_1, I_3$ and rotational angular momentum $\mathbf{L} = \boldsymbol{\mathcal{I}} \cdot \boldsymbol{\Omega}$, where $\boldsymbol{\Omega}$ is the rotational angular velocity and $\boldsymbol{\mathcal{I}}$ is the moment of inertia tensor. The magnetic moment $\boldsymbol{\mu}$ relates to the macro-spin vector $\mathbf{S}$ via the gyromagnetic ratio $\gamma = -g \frac{e}{2 m_e} = - g \frac{\mu_B}{\hbar}$ (with $g \approx 2.2$ for Co): $\boldsymbol{\mu} = \gamma \mathbf{S}$, where $e>0$ is the elementary charge, $m_e$ the electron mass, and $\mu_B$ the Bohr magneton.

When subject to external DC electric ($\mathbf{E}$) and magnetic ($\mathbf{B}$) fields, electromagnetic, relativistic motional, and spin-lattice couplings drive the system. To simulate the dynamics, conservative Hamiltonian mechanics must be paired with dissipative channels; Landau--Lifshitz--Gilbert (LLG) damping for spin relaxation \cite{LandauLifshitz1935, Gilbert2004, Gilbert1955} and mechanical friction for rigid-body rotation.

Previous treatments of magnetic-needle magnetometers and gyroscopes \cite{Kimball2016, Rusconi2017, Band2018, Wang2019, Fadeev2021, Fadeev2021b, Vinante2021, Kalia2024, Kimball2026} considered magnetic fields only via a Zeeman Hamiltonian for the macro-spin. Here we add the permanent-dipole energy, the energy of the induced electric dipole, the linear magnetoelectric coupling, the motional spin-orbit coupling, and the couplings to a static electric field, if present, i.e., the polarization torque on the conducting needle, and study their effect on the dynamics.

The paper is organized as follows: Section~\ref{sec:Hamiltonian} presents the Hamiltonian of a ferromagnetic needle placed in external magnetic and electric fields.  The polarization and the shape-dependent depolarization tensor are considered in Sec.~\ref{sec:depolarization_tensor}, and Sec.~\ref{subsec:screening} discusses how the conduction electrons in a rapidly rotating needle screen a static lab-frame electric field, why the magnetic field is not screened, and the local field at the tips.  Section~\ref{easy-axis-quaternion} parametrizes the needle easy-axis using Euler angles and quaternions.  Quaternion kinematics as the singularity-free solution is presented in Sec.~\ref{subsec:Quaternion}, Sec.~\ref{sec:fields_dissipation_angularmomentum} considers the effective magnetic field, dissipation, and total angular momentum conservation, Sec.~\ref{subsec:Gilbert} discusses the LLG dissipation term in the equations of motion for a magnetic gyro, Sec.~\ref{subsec:Joule} presents its electric analog, the Joule dissipation of the conduction electrons that carry the induced dipole, and Sec.~\ref{subsec:anisotropy-angularmomentum} deals with internal anisotropy of the magnetic gyro and the conservation of total angular momentum.  Section~\ref{sec:EofMotion} presents the equations of motion of the magnetic gyro, Sec.~\ref{subsec:EofMotionLab} writes the equations of motion in the space-fixed frame and Sec.~\ref{subsec:EofMotionBody} writes the equations of motion in the body-fixed frame.  Section~\ref{EofMDimensionless} writes the dimensionless equations of motion in both space-fixed and body-fixed frames. Section~\ref{Cobalt-parameters} lists the single-domain cobalt needle parameters, and Sec.~\ref{sec:Numerical-results} presents our numerical results for the dynamics of the magnetic gyro.  Finally, Sec.~\ref{sec:Summary} contains a summary and conclusion. 

\section{Conservative Hamiltonian Formulation}  \label{sec:Hamiltonian}

The total conservative Hamiltonian $H$ is given by the sum of the rotation, the anisotropy and the electric-magnetic Hamiltonians,
\begin{equation}  \label{eq:Hamiltonian}
H = H_{\text{rot}} + H_{\text{ani}} + H_{\text{EM}},
\end{equation}
where
\begin{align}
H_{\text{rot}} &= \frac{1}{2} \mathbf{L}^T \cdot \boldsymbol{\mathcal{I}}^{-1} \cdot \mathbf{L}, \\
H_{\text{ani}} &= -K_v V (\hat{\mathbf{m}} \cdot \hat{\mathbf{n}})^2, \\
H_{\text{EM}} &= - \boldsymbol{\mu} \cdot \mathbf{B} - \frac{\gamma}{c^2} \mathbf{S} \cdot (\mathbf{E} \times \mathbf{v}) - \mathbf{d}_0 \cdot \mathbf{E} \nonumber \\
& \quad - \frac{1}{2} \varepsilon_0 \varepsilon_m V \, \mathbf{E} \cdot \mathbf{N}_e^{-1} \cdot \mathbf{E} - \gamma \sum_{i,j} \alpha_{ij} E_i S_j.
\label{eq:HEM}
\end{align}
Here $\hat{\mathbf{m}} = \boldsymbol{\mu}/|\boldsymbol{\mu}| = - \mathbf{S}/S$ is the magnetization unit vector (so $\hat{\mathbf{m}} \equiv -\hat{\mathbf{s}}$ because $\gamma$ is negative; $\hat{\mathbf s}=\mathbf S/S$ is the normalized spin, see Sec.~\ref{EofMDimensionless}) and ${\mathbf S}$ is the total spin of the single domain needle, $\hat{\mathbf{n}}$ is the direction of the easy axis of the magnet, $K_v = K_{u1} + K_{\text{shape}}$ is the effective uniaxial anisotropy energy density given by the sum of the magnetocrystalline anisotropy constant $K_{u1}$ and the shape-anisotropy constant $K_{\text{shape}} = \frac{1}{4}\mu_0 M_s^2 (1-3N_z)$, $M_s$ is the saturation magnetization of the magnetic needle, and $N_z$ is the demagnetization factor (or depolarizing factor) along the easy axis of the magnetic needle, ${\hat {\mathbf n}}$ ($z$-axis in the body fixed frame), $V$ is the magnetic needle volume.  The electromagnetic Hamiltonian $H_{\text{EM}}$ in Eq.~(\ref{eq:HEM}), which is the interaction of the magnetic gyro with the electric and magnetic fields, is given by the sum of five contributions: the Zeeman energy $H_{\rm Z} = - \boldsymbol{\mu} \cdot {\mathbf B} = -\gamma \, \mathbf{S} \cdot \mathbf{B}$ \cite{Blundell_01, Fiebig2005, Band-Avishai-Quantum-Mechanics}, the motional spin-orbit energy, $H_{\rm SO} = - \boldsymbol{\mu} \cdot \mathbf{B}_{\text{mot}} = -\frac{\gamma}{c^2} \mathbf{S} \cdot (\mathbf{E} \times \mathbf{v})$ [see Eq.~(\ref{eq:B_mot_def}); $\mathbf{v} = \mathbf{v}_{\text{cm}} + \boldsymbol{\Omega} \times \mathbf{r}_{\text{eff}}$ is the physical velocity field, ($\mathbf{v}_{\text{cm}}$ is the center-of-mass velocity of the gyro, and $\mathbf{r}_{\text{eff}}$ is the average effective position vector given by the spin-density-weighted spatial average over the gyro volume, see Eq.~(\ref{eq:r_eff_def}) below)].  The permanent-dipole energy is given by $- \mathbf{d}_0 \cdot \mathbf{E}$, where $\mathbf{d}_0$ is the permanent electric dipole moment ($\mathbf{d}_0 = \mathbf{0}$ for inversion-symmetric materials, such as metallic cobalt).  The induced electric dipole energy is given by, $- \frac{1}{2} \varepsilon_0 \varepsilon_m V \, \mathbf{E} \cdot \mathbf{N}_e^{-1} \cdot \mathbf{E}$, where $\mathbf{N}_e^{-1} \equiv {\boldsymbol{\chi}}_{e}$ is the effective (dimensionless) polarizability tensor governed by the shape-dependent depolarization tensor $\mathbf{N}_e$, and $\varepsilon_m$ is the dielectric constant of the medium surrounding the gyroscope (for vacuum $\varepsilon_m = 1$).  Finally, the magnetoelectric energy is given by $- \gamma \sum_{i,j} \alpha_{ij} E_i S_j$, where $\alpha_{ij}$ is the linear magnetoelectric tensor (which has units of s/m).

For a spinning rigid body, the local velocity at position $\mathbf{r}$ relative to the center of mass is $\mathbf{v}(\mathbf{r}) = \mathbf{v}_{\text{cm}} + \boldsymbol{\Omega} \times \mathbf{r}$. Averaging the spin-orbit interaction over the spatial spin density distribution $\rho_s(\mathbf{r})$ (where $\int_V \rho_s(\mathbf{r}) d^3r = S$) yields an effective motional velocity coupling:
\begin{equation}
\langle \mathbf{E} \times \mathbf{v} \rangle = \mathbf{E} \times (\mathbf{v}_{\text{cm}} + \boldsymbol{\Omega} \times \mathbf{r}_{\text{eff}}),
\end{equation}
where the effective position vector $\mathbf{r}_{\text{eff}}$ is defined as the spin-density-weighted spatial average over the volume $V$:
\begin{equation} \label{eq:r_eff_def}
\mathbf{r}_{\text{eff}} \equiv \frac{1}{S} \int_V \mathbf{r} \, \rho_s(\mathbf{r}) \, d^3r.
\end{equation}
For a symmetric-top with a uniform, inversion-symmetric spin density distribution $\rho_s(\mathbf{r}) = S/V$ centered at the origin (the center of mass is taken as the origin, so $\mathbf{r}_{\text{cm}} = \mathbf{0}$), spatial symmetry dictates that $\mathbf{r}_{\text{eff}} = \mathbf{0}$. In this case, pure rotational spinning ($\mathbf{v}_{\text{cm}} = \mathbf{0}$) generates zero net volume-averaged translational spin-orbit energy. However, if the spin density distribution exhibits spatial gradients, boundary inhomogeneities, or a center-of-spin displacement relative to the mechanical pivot point ($\mathbf{r}_{\text{eff}} \neq \mathbf{0}$), the term $\boldsymbol{\Omega} \times \mathbf{r}_{\text{eff}}$ yields a non-vanishing contribution to the motional coupling $-\frac{\gamma}{c^2} \mathbf{S} \cdot [\mathbf{E} \times (\boldsymbol{\Omega} \times \mathbf{r}_{\text{eff}})]$.

\section{Polarization and Shape-Dependent Depolarization Tensor}  \label{sec:depolarization_tensor}

When a metallic ellipsoid or needle-like conductor is immersed in an external electric field $\mathbf{E}_0$, surface charges accumulate at the boundary to screen the field inside the body \cite{Jackson1999, Stratton1941, LandauLifshitzElectrodynamics}. In the body-fixed frame, and in the quasi-static limit, the geometry of the conductor is characterized by the diagonal depolarization tensor $\mathbf{N}_e = \operatorname{diag}(N_x, N_y, N_z)$, whose elements satisfy the trace condition: $\text{Tr}(\mathbf{N}_e) = N_x + N_y + N_z = 1$ \cite{Stratton1941, LandauLifshitzElectrodynamics}.  For a highly conducting metal such as cobalt ($|\varepsilon_{\text{metal}}| \gg \varepsilon_m$), interior free-charge screening causes the internal electric field to approach zero ($\mathbf{E}_{\text{in}} \to \mathbf{0}$) \cite{Jackson1999}. Consequently, the induced electric dipole moment $\mathbf{p}$ along principal axis $i \in \{x, y, z\}$ saturates to a strictly geometry-dependent bound:
\begin{equation}
\mathbf{p} = \varepsilon_0 \varepsilon_m V  \mathbf{N}_e^{-1} \cdot \mathbf{E}_0 =  \varepsilon_0 \varepsilon_m V  {\boldsymbol \chi}_e  \cdot \mathbf{E}_0,
\end{equation}
where $\boldsymbol\chi_e=\mathbf N_e^{-1}$ is the dimensionless polarizability of Eq.~(\ref{eq:HEM})].  For a prolate symmetric-top ellipsoid with semi-major axis $a$ (along the easy axis) and semi-minor axes $r$ ($a>r$) aligned along the easy axis $\hat{\mathbf{n}}$ (body-fixed $z$-axis), the depolarization factors are $N_z = N_\parallel$ and $N_x = N_y = N_\perp = \frac{1 - N_\parallel}{2}$ \cite{Stratton1941, LandauLifshitzElectrodynamics}. The shape anisotropy of the depolarization tensor produces an electrostatic polarization torque \cite{Stratton1941}:
\begin{align}  \label{eq:polarization_torque}
\mathbf{N}_{\text{elec}} &= \mathbf{p} \times \mathbf{E}_0 \nonumber \\
&= \varepsilon_0 \varepsilon_m V \left( \frac{1}{N_\parallel} - \frac{1}{N_\perp} \right) (\hat{\mathbf{n}} \cdot \mathbf{E}_0) (\hat{\mathbf{n}} \times \mathbf{E}_0).
\end{align}
The torque is $\mathbf N_{\rm elec}=-\hat{\mathbf n}\times\partial U_E/\partial\hat{\mathbf n}$ with the induced-dipole energy
\begin{align}  \label{eq:polarization_energy}
U_E &= -\frac{1}{2}\,\mathbf p\cdot\mathbf E_0=\mathrm{const}-\frac{1}{2}\,p_E\,(\hat{\mathbf n}\cdot\mathbf E_0)^2,
\end{align}
so the polarization torque tends to align the long axis of the needle with $\mathbf E_0$ (minima of $U_E$ at $\hat{\mathbf n}\parallel\pm\mathbf E_0$; the orientation $\hat{\mathbf n}\perp\mathbf E_0$ is an unstable maximum). It is governed by the scalar polarizability anisotropy (which has SI units of C\,m$^2$/V)
\begin{align}
p_E = \varepsilon_0 \varepsilon_m V \left( \frac{1}{N_\parallel} - \frac{1}{N_\perp} \right). \label{eq:pE}
\end{align}
The orientation-dependent part of $U_E$, of magnitude $\tfrac{1}{2} p_E E_0^2$, competes with the Zeeman energy $-\boldsymbol\mu\cdot\mathbf B$, of magnitude $\mu B$ with $\mu=M_sV$; correspondingly, the maximum polarization torque, $\tfrac{1}{2}p_EE_0^2$ [at $\angle(\hat{\mathbf n},\mathbf E_0)=\pi/4$], is compared with the maximum magnetic torque, $\mu B$. We define the crossover field $E_c$ by $p_EE_c^2=\mu B$:
\begin{equation}\label{eq:Ec}
E_c=\left[\frac{M_sB}{\varepsilon_0\varepsilon_m\,(N_\parallel^{-1}-N_\perp^{-1})}\right]^{1/2}.
\end{equation}
With this convention the maximum polarization torque at $E_c$, $\tfrac{1}{2} p_EE_c^2$ is half the maximum magnetic torque $\mu B$; the two maxima are equal at $\sqrt2 E_c \approx 1.2 \times 10^{8}$~V/m at 0.5~T. For $E_0\ll E_c$ the magnetic torque dominates the orientation of the needle, and for $E_0\gg E_c$ the polarization torque dominates. Since $p_E$ and $\mu$ are both proportional to $V$, $E_c$ is independent of the size of the needle; it depends on its aspect ratio through $N_\parallel$ and $N_\perp$. For the cobalt needle of Sec.~\ref{Cobalt-parameters}, $E_c\approx8.5\times10^{7}$~V/m at $B=0.5$~T and $E_c\approx2.7\times10^{5}$~V/m at $B=5~\mu$T.

\subsection{Electrostatic screening of a cobalt nanoneedle}
\label{subsec:screening}

In the body-fixed frame the static lab-frame field $\bm E$ appears as
$\bm E_b(t)={\bm R}^{T}(t)\,\bm E$, whose component transverse to the rotation
axis rotates at the angular frequency ${\boldsymbol \Omega}$. Here we ask whether the
conduction electrons of a cobalt needle, modeled as a prolate spheroid with
semi-axes $a=60$~nm and $r=15$~nm (depolarization factors $N_\parallel=0.075$
and $N_\perp=0.46$, tip radius of curvature $r_{\rm tip}=r^{2}/a=3.75$~nm),
can screen this time-dependent field as effectively as a static one, for
rotation rates $\Omega/2\pi\sim1$~GHz. We then examine the local field at the
tips of the needle.

\paragraph{Electric field: screening persists.}
For $\Omega \tau_p \ll 1$, where $\tau_p$ is the momentum-relaxation time (not to be confused with the dimensionless time $\tau$ of Sec.~\ref{EofMDimensionless}), which in the Drude model \cite{AshcroftMermin1976} is $\tau_p = m_e \sigma/(n_e e^2)$ with $n_e$ the conduction-electron density, $m_e$ the electron mass, $e$ the elementary charge, and $\sigma$ the dc conductivity [for Co, with $\sigma \approx 1.6\times10^{7}$~S/m and about one conduction electron per atom, $n_e\approx9\times10^{28}$~m$^{-3}$, this gives $\tau_p \approx 6 \times 10^{-15}$~s, so that $\Omega \, \tau_p \approx 4 \times 10^{-5}$ at $\Omega/2\pi=1$~GHz], the
dielectric function of a good conductor is dominated by its imaginary part, $\varepsilon(\Omega)\simeq 1+i\sigma/(\varepsilon_0\Omega)\approx i\sigma/(\varepsilon_0\Omega)$ [this follows from the Drude form, $\varepsilon(\Omega)=1-\omega_p^2/[\Omega(\Omega+i/\tau_p)]$, with plasma frequency $\omega_p=(n_ee^2/\varepsilon_0m_e)^{1/2}\approx1.7\times10^{16}$~s$^{-1}$: for $\Omega \tau_p \ll 1$ and $\Omega\ll\omega_p$, $\varepsilon\simeq i\omega_p^2 \tau_p/\Omega = i \sigma/(\varepsilon_0\Omega)$, because $\sigma=\varepsilon_0\omega_p^2 \tau_p$]. The room-temperature
dc conductivity of bulk cobalt is $\sigma \approx 1.6\times 10^{7}$~S/m, which gives
\begin{equation}
|\varepsilon| \simeq \frac{\sigma}{\varepsilon_0\Omega} \approx 3 \times 10^{8}
\qquad (\Omega/2\pi = 1~\mathrm{GHz}),
\end{equation}
and a charge-relaxation time $\tau_\chi \equiv \varepsilon_0/\sigma \approx 5\times10^{-19}$~s. The Thomas--Fermi screening length $\lambda_{\rm TF}$ is a fraction of a nanometer.
All of these are many orders of magnitude shorter than $2\pi/\Omega$ and than
the particle size. In the quasi-static limit the polarizability of a spheroid
along principal axis $i$ (for $\varepsilon_m = 1$) [in the perfect-conductor limit it is $\chi_i(\varepsilon) = \varepsilon_0 \varepsilon_m V\, \chi_{e,ii}(\varepsilon)$; in SI units, C\,m$^2$/V] is
\begin{equation}  \label{eq:polarizability}
\chi_i(\varepsilon) = \varepsilon_0 V \, \frac{\varepsilon-1}{1+N_i(\varepsilon-1)}
\simeq\frac{\varepsilon_0V}{N_i}\Big[1-\frac{1}{N_i\,\varepsilon}\Big].
\end{equation}
The deviation from the perfect-conductor value $\varepsilon_0V/N_i$ is
$\sim1/(N_i|\varepsilon|)\lesssim5\times10^{-8}$, and it is almost entirely
out of phase with the field. The induced dipole therefore follows $\bm E_b(t)$
with a phase lag of order $10^{-8}$~rad. The field inside the metal remains zero
to excellent accuracy, the electric torque of Eq.~(\ref{eq:polarization_torque}) holds with $\bm E\to\bm E_b(t)$ and $\chi_i=\varepsilon_0V/N_i$, and the dissipative
part of the electric torque (derived in Sec.~\ref{subsec:Joule}) is smaller than the conservative part by
$\sim10^{-8}$. The same conclusion holds if $\sigma$ is reduced by several
orders of magnitude by surface scattering in a nanoparticle. Screening would fail only for $\Omega \tau_p \gtrsim 1$ or near the surface-plasmon frequency, at
THz-to-optical frequencies, orders of magnitude above the rotation rates
considered here.

\paragraph{Local field at the tips (lightning-rod effect).}
The electric field is not uniform outside the needle.  For a conducting ellipsoid in a uniform field, the surface charge density is $\sigma_s = \varepsilon_0\sum_iE_{b,i}\,\nu_i/N_i$, where $\bm\nu$ is the outward unit normal, so the normal field at the surface is
\begin{equation}
E_n(\bm\nu)=\sum_{i=1}^{3}\frac{E_{b,i}(t)}{N_i}\,\nu_i .
\end{equation}
At the tip ($\bm\nu=\pm\hat{\bm n}$) this gives
\begin{equation}
E_{\rm tip}(t)=\frac{E_{b3}(t)}{N_\parallel}=\frac{\bm E\!\cdot\!\bm n(t)}{N_\parallel}
\approx13.3\,\bm E\!\cdot\!\bm n(t),
\end{equation}
while at the equator the enhancement is only $1/N_\perp\approx2.2$. The tip
field is therefore enhanced by an order of magnitude and oscillates at the rotation
frequency, since $\bm E\!\cdot\!\bm n$ varies as the needle rotates. Its
maximum, $E_0/N_\parallel$, is reached whenever $\bm n\parallel\bm E$. In the simulations of Sec.~\ref{sec:Numerical-results} (Fig.~\ref{EB_low.needle_easy_axis} for the magnetic gyro in a strong electric field of $E_0=5\times 10^{7}$~V/m along the $x$-axis) the needle swings into alignment with $\bm E$ [$|\bm E\!\cdot\!\bm n|/E_0$ reaches values very close to 1], so this maximum is attained.  For the field used in Fig.~\ref{EB_low.needle_easy_axis}, the electric field at the tip reaches a maximum value of $E_{\rm tip} \approx 0.66$~V/nm every rotational period.

\paragraph{Effect on the rotational dynamics.}
The lightning-rod effect does \emph{not} change the electric torque of Eq.~(\ref{eq:polarization_torque}) or the energy $U_E$ of Eq.~(\ref{eq:polarization_energy}). These are determined by the total induced dipole $\bm p=\oint\sigma_s\bm r\,dA= \bm \chi \bm E_b$, and the large charge density at the tips is precisely what produces the large $\chi_\parallel=\varepsilon_0V/N_\parallel$ and hence the anisotropy $\Delta\chi \equiv \chi_\parallel-\chi_\perp$ $=p_E$ [Eq.~(\ref{eq:pE})]. The enhancement is therefore already contained in the polarizability. It matters only through the local conditions at the surface, discussed next. The charge sloshing between the tips at frequency $\Omega$ amounts to $\sim 20$ electrons in the field of $E_0=5\times 10^{7}$~V/m (the oscillating tip charge is $Q \simeq p/2a$). It drives a current $\sim \Omega \, Q \sim 2 \times 10^{-8}$~A through a needle of resistance $\sim 10~{\rm ohm}$, which dissipates an energy per cycle $\sim10^{-7}$ of the
electric energy $\tfrac{1}{2} \Delta\chi E_0^2$, consistent with the out-of-phase
fraction $1/(N_\parallel|\varepsilon|)$ estimated above.

\paragraph{Validity of the model at the tip.}
The large local field limits the applicability of the induced-dipole model in
four ways.
\begin{enumerate}
\item \emph{Classical electrostatics.} The tip radius $r_{\rm tip}=3.75$~nm is much
larger than $\lambda_{\rm TF}\sim0.1$~nm and the Fermi wavelength, so the local
classical treatment holds. Nonlocal and electron-spill-out corrections to the
effective polarizability are of relative order $\lambda_{\rm TF}/r_{\rm tip}$, a few percent.
The surface charge at $E_0=5\times10^{7}$~V/m, $\sigma_s = \varepsilon_0E_{\rm tip} \approx 0.006$~C/m$^2$ ($\approx 0.04$ electrons/nm$^2$), is only $\sim 0.2\%$ of an electron per surface atom, so the screening layer is not saturated.
\item \emph{Field emission.} For a work function $\Phi\approx5$~eV (a typical value
for cobalt), the Fowler--Nordheim current density is
$J\simeq(a_{\rm FN} F^{2}/\Phi)\exp(-b_{\rm FN} \Phi^{3/2}/F)$ (image-charge correction neglected) with $a_{\rm FN} = 1.54 \times 10^{-6}$~A  eV V$^{-2}$, $b_{\rm FN} = 6.83$~V eV$^{-3/2}$\,nm$^{-1}$ and $F=E_{\rm tip}$ ($B$ is reserved for the magnetic field). For an emitting area of $\sim10$~nm$^{2}$ this gives $\sim10^{3}$, $10^{6}$, $10^{8}$ and $10^{10}$ electrons/s at $F=3$, $4$, $5$ and $6.6$~V/nm, respectively. Emission is thus negligible for $F \lesssim 3$~V/nm, that is, for $E_0 \lesssim 2 \times 10^{8}$~V/m. Emitted electrons would charge the particle, which would add a net force and a monopole contribution that the induced-dipole model does not contain. \item \emph{Electrostatic stress.} The Maxwell stress at the tip is $\varepsilon_0E_{\rm tip}^{2}/2 \approx 2\times10^{6}$~Pa at $E_0=5\times10^{7}$~V/m, far below the bulk tensile strength of cobalt (a few hundred MPa); since the stress scales as $E_0^2$, it would reach this limit at $E_0\approx5\times10^{8}$~V/m. 
\item \emph{Field evaporation.} Atom-probe evaporation fields for transition metals are typically several tens of V/nm, well above the fields considered here.
\end{enumerate}
The first two conditions restrict the lab-frame field to $E_0 \lesssim 2 \times 10^{8}$~V/m
for the needle considered here, and more strongly for sharper tips (smaller
$N_\parallel$).

\paragraph{Magnetic field: not screened.}
The skin depth is
\begin{equation}
\delta=\sqrt{\frac{2}{\mu_0\mu_r\sigma\Omega}}\approx\frac{4~\mu\mathrm{m}}{\sqrt{\mu_r}}
\qquad(\Omega/2\pi=1~\mathrm{GHz}).
\end{equation}
Even for an effective permeability $\mu_r\sim100$ it is $\sim0.4~\mu$m, which
exceeds the needle dimensions ($\sim120$~nm long and $\sim30$~nm across), so
eddy currents do not screen the field and the full magnetic field acts on the
magnetization. Because the needle is a single-domain particle whose
magnetization follows the Landau--Lifshitz--Gilbert equation, the bulk
permeability should not be used to rescale the field inside it. The time
dependence of $\bm B_b(t)$ in the body frame drives eddy currents of density
$j\sim\sigma\Omega rB$, which add a viscous rotational damping coefficient $\Gamma_{\rm eddy}\sim \sigma \, B^2 V \, r^2 \sim 5 \times 10^{-32}$~N\,m\,s for $B\sim0.5$~T. This is negligible: even at $\Omega=10^{9}$~rad/s the resulting torque is $\sim5\times10^{-23}$~N\,m, against an electric torque of order $\Delta\chi E_0^2 \sim 10^{-17}$~N\,m at $E_0 = 5 \times 10^{7}$~V/m.

\section{Easy-Axis Euler Parameterization and Quaternions}  \label{easy-axis-quaternion}

The spatial orientation of a symmetric-top magnetic needle is characterized by the direction of its easy magnetization axis $\hat{\mathbf{n}}(t)$. In the laboratory frame, $\hat{\mathbf{n}}$ is parameterized by spherical polar angle $\theta(t) \in [0, \pi]$ (nutation angle) and azimuthal angle $\phi(t) \in [0, 2\pi)$ (precession angle):
\begin{equation}
\hat{\mathbf{n}}(\theta, \phi) = \begin{pmatrix} \sin\theta \cos\phi \\ \sin\theta \sin\phi \\ \cos\theta \end{pmatrix}.
\end{equation}
To map laboratory basis vectors $(\hat{\mathbf{e}}_x, \hat{\mathbf{e}}_y, \hat{\mathbf{e}}_z)$ to the body-fixed basis $(\hat{\mathbf{e}}_{1,b}, \hat{\mathbf{e}}_{2,b}, \hat{\mathbf{e}}_{3,b} \equiv \hat{\mathbf{n}})$, we apply the standard active Euler rotation sequence $\mathbf{R}(\phi, \theta, \psi) = \mathbf{R}_z(\phi) \mathbf{R}_y(\theta) \mathbf{R}_z(\psi)$ ($z$-$y$-$z$ convention), where $\psi(t)$ represents intrinsic body rotation about the easy axis $\hat{\mathbf{n}}$.

Explicitly, the full active 3-degree-of-freedom rotation matrix $\mathbf{R}(\theta, \phi, \psi) \in SO(3)$ is:
\begin{widetext}
\begin{equation}
\mathbf{R}(\theta, \phi, \psi) = \begin{pmatrix}
\cos\phi\cos\theta\cos\psi - \sin\phi\sin\psi & -\cos\phi\cos\theta\sin\psi - \sin\phi\cos\psi & \sin\theta\cos\phi \\
\sin\phi\cos\theta\cos\psi + \cos\phi\sin\psi & -\sin\phi\cos\theta\sin\psi + \cos\phi\cos\psi & \sin\theta\sin\phi \\
-\sin\theta\cos\psi & \sin\theta\sin\psi & \cos\theta
\end{pmatrix}.
\label{eq:R_full_euler}
\end{equation}
\end{widetext}

Although specifying $\hat{\mathbf{n}}(\theta, \phi)$ fixes the orientation of the 3rd body axis, fixing or setting $\psi = 0$ is insufficient to carry out full rigid-body dynamics. The intrinsic angle $\psi(t)$ evolves according to the body-frame rotational angular velocity component $\Omega_{b3}(t) = \boldsymbol{\Omega} \cdot \hat{\mathbf{n}}$. For the axisymmetric needle considered here ($I_1=I_2$, $\alpha_{xx}=\alpha_{yy}$) the laboratory-frame tensors $\boldsymbol{\mathcal I}(t)=I_1{\mathbf 1}+(I_3-I_1)\hat{\mathbf n}\hat{\mathbf n}^T$ and $\boldsymbol\alpha(t) = \alpha_\perp{\mathbf 1}+(\alpha_\parallel-\alpha_\perp)\hat{\mathbf n}\hat{\mathbf n}^T$ depend only on $\hat{\mathbf n}$, so $\psi$ is a cyclic coordinate and $(\hat{\mathbf n},\mathbf L,\mathbf S)$ obey a closed set of equations without it. Tracking $\psi(t)$ (or the full rotation matrix $\mathbf R$) is needed only for the following purposes:
\begin{enumerate}
    \item \textbf{Body-frame description:}  The body-frame description of Sec.~\ref{subsec:EofMotionBody}, in which $\boldsymbol{\Omega}_b=\boldsymbol{\mathcal I}_0^{-1}\mathbf L_b$ and $\mathbf E_b$, $\mathbf B_b$ are obtained with the full $\mathbf R(t)$. In the laboratory frame $\boldsymbol{\Omega}=\mathbf L/I_1+(I_3^{-1}-I_1^{-1})(\hat{\mathbf n}\!\cdot\!\mathbf L)\,\hat{\mathbf n}$ requires no $\psi$.
    \item \textbf{Body-frame components and symmetry-breaking extensions:} The individual body-frame components $S_{b1},S_{b2}$ depend on $\psi$ (physical observables do not), and any extension that breaks axial symmetry (triaxial shape, hexagonal basal-plane anisotropy of Co, or an anisotropic $\boldsymbol\alpha_\perp$) requires $\psi(t)$.
\end{enumerate}

A common approach to tracking the easy axis vector $\hat{\mathbf{n}}(t) = (n_x(t), n_y(t), n_z(t))^T$ is to extract Euler angles dynamically using standard trigonometric relations:
\begin{align}
\theta(t) &= \arccos\left(\frac{n_z(t)}{\sqrt{n_x(t)^2 + n_y(t)^2 + n_z(t)^2}}\right), \label{eq:theta_def} \\
\phi(t) &= \arctan\left(\frac{n_y(t)}{n_x(t)}\right). \label{eq:phi_def}
\end{align}
While mathematically valid for isolated static configurations, employing spherical angles inside differential equation solvers introduces severe numerical pathologies:

\begin{enumerate}
    \item \textbf{Gimbal Lock at the Polar Boundaries ($\theta \to 0, \pi$):}
    As the needle precesses near the vertical lab axes ($n_x \to 0, n_y \to 0$), the expression $\arctan(n_y / n_x)$ approaches the indeterminate form $0/0$. Physically, when the nutation angle $\theta = 0$, the azimuthal angle $\phi$ and intrinsic spin $\psi$ become collinear and linearly dependent, losing one rotational degree of freedom.
    \item \textbf{Divergence of Angular Velocities and Step-Size Failure:}
    The kinematic transformation relating dimensionless rotational angular velocity $\tilde{\boldsymbol{\Omega}}$ to Euler angle rates $(\dot{\phi}, \dot{\theta}, \dot{\psi})$ contains a factor of $1/\sin\theta$:
    \begin{align}
    \dot\phi &= \frac{\tilde\Omega_{b2}\sin\psi - \tilde\Omega_{b1}\cos\psi}{\sin\theta}\quad 
    (z\text{-}y\text{-}z),\\
    \dot\theta &= \tilde\Omega_{b1}\sin\psi + \tilde\Omega_{b2}\cos\psi,\quad
    \dot\psi = \tilde\Omega_{b3} - \dot\phi\cos\theta .
    \end{align}
As $\theta \to 0$ or $\theta \to \pi$, $\sin\theta \to 0$, causing time derivatives $\dot{\phi}(t)$ to diverge to infinity ($\dot{\phi} \to \infty$). Consequently, adaptive step-size algorithms would reduce the time step $\Delta t \to 0$ to maintain error tolerances, ultimately leading to integration halts or catastrophic loss of precision.  The quaternion formulation below removes the coordinate singularity altogether.

    \item \textbf{Branch Cut Discontinuities:}
    Evaluating $\arctan(n_y/n_x)$ introduces step-discontinuities of $\pm \pi$ as $n_x$ crosses zero. Even using the two-argument \texttt{ArcTan[nx, ny]}, derivative evaluations within implicit numerical schemes trigger stiffness artifacts across quadrant boundaries.
\end{enumerate}

\subsection{Quaternion Kinematics as the Singularity-Free Solution}  \label{subsec:Quaternion}

To resolve coordinate singularities completely in either the space-fixed or body-fixed reference frame, orientation should be parameterized using normalized quaternions $\mathbf{q}(t) = (q_0, q_1, q_2, q_3)^T$ with unit norm condition $|\mathbf{q}|^2 = 1$ (scalar-first convention; the norm is preserved by the exact flow and should be enforced numerically by renormalization or a projection method).

The continuous transformation matrix $\mathbf{R}(\mathbf{q})$ mapping body-fixed coordinates to laboratory coordinates is given quadratically by Eq.~\eqref{eq:rot_matrix_quaternions} below. Because the dynamic rate equations for $\mathbf{q}(t)$ contain no trigonometric functions, divisions, or square roots, they remain smooth, continuous, and singularity-free across all orientations (they are parametrized by $S^3\to SO(3)$), enabling a fast and stable method of tracking the easy axis dynamics in the space-fixed representation of the dynamics or the rotation of the electric and magnetic fields in the body fixed representation.

To extract the $z$--$y$--$z$ Euler angles of Eq.~(\ref{eq:R_full_euler}), i.e., the angles used in Eqs.~(\ref{eq:theta_def}) and (\ref{eq:phi_def}) and in the kinematic relations above, from a given unit quaternion $(q_0, q_1, q_2, q_3)$ one uses \cite{Bernardes2022, Kuipers1999} (Mathematica's \texttt{EulerMatrix[\{$\phi,\theta,\psi$\}]} and \texttt{EulerAngles} use this $z$--$y$--$z$ sequence by default; the formulas for the $z$--$x$--$z$ sequence are different and must not be mixed with those of Eq.~(\ref{eq:R_full_euler})):
\begin{equation} \label{eq:Euler-angles-dynamics}
\begin{aligned}
\phi(t) &= \operatorname{atan2}(R_{23},R_{13})\\
&= \operatorname{atan2}\big(q_2 q_3 - q_0 q_1,\ q_1 q_3 + q_0 q_2\big),\\
\theta(t) &= \arccos(R_{33}) = \arccos\big(1 - 2(q_1^2 + q_2^2)\big),\\
\psi(t) &= \operatorname{atan2}(R_{32},-R_{31})\\
&= \operatorname{atan2}\big(q_2 q_3 + q_0 q_1,\ q_0 q_2 - q_1 q_3\big).
\end{aligned}
\end{equation}
These follow from the third column and the third row of Eq.~(\ref{eq:R_full_euler}), $R_{13}=\sin\theta\cos\phi$, $R_{23}= \sin\theta \sin \phi$, $R_{33} = \cos\theta$, $R_{31} = -\sin\theta\cos\psi$, $R_{32}= \sin\theta \sin\psi$, together with Eq.~(\ref{eq:rot_matrix_quaternions}); they are defined for $\sin\theta \neq 0$ (at $\theta=0,\pi$ only $\phi+\psi$, respectively $\phi-\psi$, is determined). For the initial rotation by $\pi/6$ about the $x$ axis used in Sec.~\ref{sec:Numerical-results}, $\mathbf q(0) = (\cos\frac{\pi}{12}, \sin\frac{\pi}{12},0,0)^T$, they give $\theta=\pi/6$, $\phi = -\pi/2$, $\psi = \pi/2$.

\section{Effective Fields, Dissipation, and Total Angular Momentum Conservation}  \label{sec:fields_dissipation_angularmomentum}

\subsection{Gilbert Dissipation}  \label{subsec:Gilbert}

The effective magnetic field $\mathbf{B}_{\text{eff}}$ experienced by the macro-spin is obtained from the gradient of the Hamiltonian with respect to $\boldsymbol\mu$ \cite{Gilbert2004}:
\begin{equation}  \label{eq:B_eff}
\mathbf{B}_{\text{eff}} =
- \frac{\partial H}{\partial \boldsymbol{\mu}} =
\mathbf{B} + \frac{2 K_v}{M_s} (\hat{\mathbf{m}} \cdot \hat{\mathbf{n}}) \hat{\mathbf{n}} +
\mathbf{B}_{\text{ME}} + \mathbf{B}_{\text{mot}},
\end{equation}
where 
\begin{align}  \label{eq:magnetoelectric}
(\mathbf{B}_{\text{ME}})_i = \sum_j \alpha_{ji} E_j,
\end{align}
is the induced magnetic field proportional to the product of the linear magnetoelectric tensor $\alpha_{ij}$ and the electric field $E_j$. To first order in $v/c$ relativistic motional magnetic field $\mathbf{B}_{\text{mot}}$ experienced by a spin moving with velocity $\mathbf v$ through a static field $\mathbf E$ is given by:
\begin{equation} \label{eq:B_mot_def}
\mathbf{B}_{\text{mot}} = \frac{1}{c^2} \mathbf{E} \times \mathbf{v} .
\end{equation}
Here $\mathbf{v} = \mathbf{v}_{\text{cm}} + \boldsymbol{\Omega} \times \mathbf{r}_{\text{eff}}$ is the effective velocity of the spin distribution relative to the laboratory frame. As defined in Eq.~\eqref{eq:r_eff_def}, the effective radius $\mathbf{r}_{\text{eff}} = \frac{1}{S} \int_V \mathbf{r} \rho_s(\mathbf{r}) d^3r$ accounts for the spatial distribution of the macro-spin within the rotating body. For a spinning symmetric-top whose spin density is centered at the origin ($\mathbf{r}_{\text{cm}} = \mathbf{0}$ and $\mathbf{r}_{\text{eff}} = \mathbf{0}$), the rotational velocity contribution vanishes ($\boldsymbol{\Omega} \times \mathbf{r}_{\text{eff}} = \mathbf{0}$), reducing $\mathbf{v}$ simply to the center-of-mass translation $\mathbf{v}_{\text{cm}}$. Conversely, for an off-center or asymmetrical spin distribution, $\mathbf{r}_{\text{eff}}$ provides the effective lever arm through which $\boldsymbol{\Omega}$ produces an additional velocity component that contributes to $\mathbf{B}_{\text{mot}}$.  If the spin density is not centered at the origin, e.g., only the top-half of the gyro is magnetic, or, if the rotation of the gyro is not around the center of mass, then $\mathbf{r}_{\text{eff}}$ does not vanish, and therefore the $ \boldsymbol{\Omega} \times \mathbf{r}_{\text{eff}}$ contribution to $\mathbf{B}_{\text{mot}}$ does not vanish.

Dissipative dynamics are incorporated via the LLG dissipation function density $\mathcal{R}_G$ \cite{Gilbert2004, Band2018}:
\begin{equation}
\mathcal{R}_G = \frac{\alpha_G}{2 |\gamma| M_s} \left| \frac{d\mathbf{M}}{dt} - \boldsymbol{\Omega} \times \mathbf{M} \right|^2,
\end{equation}
where $\alpha_G$ is the dimensionless LLG damping parameter and $\mathbf{M} = \gamma \mathbf{S} / V$. The frame-covariant time derivative $\frac{d\mathbf{M}}{dt} - \boldsymbol{\Omega} \times \mathbf{M}$ maintains exact frame-indifference under mechanical rotation \cite{Chudnovsky2005, Band2018}. The Gilbert equation follows from $\dot{\mathbf M}=-|\gamma|\,\mathbf M\times(\mathbf B_{\rm eff}+\mathbf B_{\rm diss})$ with the dissipative field $\mathbf B_{\rm diss}=-\partial\mathcal R_G/\partial\dot{\mathbf M}$ (with $\dot{\mathbf M}\to\dot{\mathbf M}-\boldsymbol\Omega\times\mathbf M$ in the rotating frame), which gives Eq.~(\ref{eq:LLG.Gilbert}) below; the reaction torque on the magnetic gyro follows from the dependence of $\mathcal R_G$ on $\boldsymbol\Omega$ and is equal and opposite to the damping torque on the spin.

The dissipative dynamics of a macrospin with unit magnetization $\mathbf m=\mathbf M/M_s$ ($|\mathbf m|=1$)
in an effective field $\mathbf B_{\text{eff}}$ is described by the Landau--Lifshitz--Gilbert (LLG) equation
\begin{equation}    \label{eq:LLG.Gilbert}
   \dot{\mathbf m} = -|\gamma| \,\mathbf m\times\mathbf B_{\text{eff}}+\alpha_G\,\mathbf m\times\dot{\mathbf m}.
\end{equation}
Solving Eq.~(\ref{eq:LLG.Gilbert}) for $\dot{\mathbf m}$ [take the cross product of Eq.~(\ref{eq:LLG.Gilbert}) with $\mathbf m$, use $\mathbf m\cdot\dot{\mathbf m}=0$, and substitute $\mathbf m\times\dot{\mathbf m}$ back], the Gilbert equation is brought to the equivalent Landau--Lifshitz form
\begin{equation}     \label{eq:LLG.LL}
   \dot{\mathbf m}=-\frac{|\gamma|}{1+\alpha_G^2}\,\mathbf m\times\mathbf B_{\text{eff}}
   -\frac{\alpha_G|\gamma|}{1+\alpha_G^2}\,\mathbf m\times(\mathbf m\times\mathbf B_{\text{eff}}).
\end{equation}
In polar coordinates, $\mathbf m=(\sin\theta\cos\phi,\sin\theta\sin\phi,\cos\theta)$, where $\theta$ and $\phi$ are the polar and azimuthal angles of $\mathbf m$ about the direction of $\mathbf B_{\rm eff}$ (taken as $\hat{\mathbf z}$) and are not the Euler angles of Sec.~\ref{easy-axis-quaternion}, with $\mathbf m\times\hat{\mathbf z}=-\sin\theta\,\hat{\boldsymbol\phi}$ and
$\mathbf m\times(\mathbf m\times\hat{\mathbf z})=\cos\theta\,\mathbf m-\hat{\mathbf z}$, whose component along $\hat{\boldsymbol\theta}$ is $\sin\theta$, Eq.~(\ref{eq:LLG.LL}) gives
\begin{equation}    \label{eq:LLG.polar}
   \dot\phi=\omega_B\equiv\frac{|\gamma| B_{\text{eff}}}{1+\alpha_G^2},\qquad
   \dot\theta=-\alpha_G\,\omega_B\,\sin\theta .
\end{equation}
The magnetization precesses around the field at the Larmor angular frequency (corrected by the LLG parameter) $\omega_B$ while its polar angle relaxes toward the field direction. The second of
Eqs.~(\ref{eq:LLG.polar}) integrates to $\tan(\theta/2)=\tan(\theta_0/2)\,e^{-\alpha_G \omega_B t}$, so that the transverse components $m_{x,y}=\sin\theta\,(\cos\phi,\sin\phi)$ decay for small $\theta$ as $e^{-t/\tau_G}$ with the LLG damping time
\begin{equation}    \label{eq:LLG.tau}
   \tau_G=\frac{1}{\alpha_G\,\omega_B}=\frac{1+\alpha_G^2}{\alpha_G\,|\gamma| B_{\text{eff}}} \simeq \frac{1}{\alpha_G\,|\gamma| B_{\text{eff}}} \quad (\alpha_G\ll1).
\end{equation}
Thus the damping time is the inverse of the precession frequency multiplied by $\alpha_G$: the magnetization performs about $1/(2\pi\alpha_G)$ precession cycles before the transverse amplitude drops by a factor $e$. The same time governs the approach of $m_z=\cos\theta$ to $1$ for small $\theta$, whereas the energy $\mathcal E = -M_s V\,\mathbf m\cdot\mathbf B$ of a sample of volume $V$ relaxes, near the minimum, at the rate $2/\tau_G$ because ${\mathcal E}-{\mathcal E}_{\min}\propto\theta^2$. For example, for the cobalt needle of Sec.~\ref{Cobalt-parameters} the effective field is dominated by the anisotropy field $2K_v/M_s\approx1.3$~T; with $B=0.5$~T ($|\mathbf B_{\rm eff}(0)| \approx 0.88$~T) and $\alpha_G = 0.01$ this gives $\tau_G \approx 0.6$~ns.  Here, $\tau_G$ was calculated using the initial value of the effective magnetic field, but in our calculation the effective magnetic field $B_{\rm eff}$ oscillates rapidly with the Larmor frequency.  Therefore, in the calculations the effective Gilbert damping time is actually roughly 0.6 ns in Fig.~\ref{Fig_magnetic_needle_spin} where the magnetic field strength was taken to be 0.5 T.

\subsection{Electric Analog of Gilbert Damping: Joule Dissipation of the Induced Dipole} \label{subsec:Joule}

The induced electric dipole of the conducting needle also has dissipative dynamics, due to the conduction electrons that carry the polarization charge. It is the electric analog of Gilbert damping, but it is not of the Landau--Lifshitz--Gilbert form: the dipole moment $\mathbf p$ carries no intrinsic angular momentum, so its equation of motion has no precession term, and its length is not conserved. The dynamics is instead a Debye relaxation \cite{Debye1929} of $\mathbf p$ toward its instantaneous equilibrium value.

\paragraph{Debye relaxation of the induced dipole.} At frequencies $\omega\ll1/\tau_p$, where $\tau_p$ is the Drude momentum-relaxation time of Sec.~\ref{subsec:screening} and $\tau_{\chi} \equiv \frac{\varepsilon_0}{\sigma}$ is the charge-relaxation time, the Drude dielectric function of the metal is $\varepsilon(\omega) \simeq 1+i\sigma/(\varepsilon_0\omega) = 1 + \tfrac{i}{\tau_{\chi} \omega}$ (Sec.~\ref{subsec:screening}) \cite{AshcroftMermin1976}. Substituting this form into the quasi-static polarizability of a spheroid along principal axis $i$ (for $\varepsilon_m=1$)  \cite{BohrenHuffman1983, LandauLifshitzElectrodynamics}, see Eq.~(\ref{eq:polarizability}), gives
\begin{align}
\chi_i(\omega)&=\varepsilon_0 V\,\frac{\varepsilon(\omega)-1}{1+N_i(\varepsilon(\omega)-1)}
=\frac{\varepsilon_0V}{N_i-i\varepsilon_0\omega/\sigma} \nonumber \\
&=\frac{\chi_i}{1-i\omega\tau_{\chi,i}},
\label{eq:tau_i}
\end{align}
with
\begin{align}
\tau_{\chi,i} &= \frac{\varepsilon_0}{N_i\,\sigma} = \frac{\tau_{\chi}}{N_i}
\end{align}
the decay time for the $i$ component of the induced dipole, and $\chi_i=\varepsilon_0 V /N_i$ the static (perfect-conductor) polarizability and $e^{-i\omega t}$ time dependence. This is a Debye response. In the body (lattice) frame, where the polarizability tensor is diagonal and constant, the induced dipole therefore obeys the driven differential equation,
\begin{equation} \label{eq:Debye_body}
\tau_{\chi,i}\left(\frac{d\mathbf p_b}{dt}\right)_{\!{\rm body},i}+p_{b,i}=\chi_i\,E_{b,i}(t),
\quad \mathbf E_b(t)=\mathbf R^T(t)\mathbf E,
\end{equation}
and in the laboratory frame $(d\mathbf p/dt)_{\rm body}\to d\mathbf p/dt -\boldsymbol\Omega \times \mathbf p$, the same frame-covariant derivative that appears in the Gilbert dissipation function. If the inertia of the electrons is retained (finite $\tau_p$), Eq.~(\ref{eq:Debye_body}) is replaced by the equation of a damped oscillator at the surface-plasmon frequency $\omega_i=N_i^{1/2}\omega_p$,
$$\ddot p_{b,i} + \tau_p^{-1}\dot p_{b,i} + \omega_i^2 p_{b,i} = \omega_i^2\chi_i E_{b,i}.$$
For driving frequencies $\omega\ll\omega_i$ the inertial term is negligible and the response reduces to Eq.~(\ref{eq:Debye_body}) with $\tau_{\chi,i}=1/(\omega_i^2 \tau_p) = \varepsilon_0/(N_i\sigma)$, since $\sigma = \varepsilon_0\omega_p^2 \tau_p$. [The surface plasmon itself is underdamped, $\omega_i\tau_p\gg1$, but this is irrelevant at the rotation frequencies, $\Omega\ll\omega_i$.]

\paragraph{Dissipation function.} Equation~(\ref{eq:Debye_body}) follows from $\partial U_p/\partial\mathbf p + \partial\mathcal R_E/\partial\dot{\mathbf p}= \mathbf 0$, where $\mathcal R_E$ is a Rayleigh dissipation function \cite{Goldstein2002}, with the energy of the induced dipole $U_p=\sum_ip_{b,i}^2/(2\chi_i)-\mathbf p\cdot\mathbf E$ and the dissipation function
\begin{equation} \label{eq:R_E}
\mathcal R_E=\frac{1}{2V\sigma}\left|\frac{d\mathbf p}{dt}-\boldsymbol\Omega\times\mathbf p\right|^2 ,
\end{equation}
because $\tau_{\chi,i}/\chi_i=1/(V\sigma)$ for every $i$. Thus $\mathcal R_E$ is isotropic although the polarizability is not, and it has the same structure as the Gilbert dissipation function $\mathcal R_G=(\alpha_G/2|\gamma|M_s)|d\mathbf M/dt-\boldsymbol\Omega\times\mathbf M|^2$, with $\mathbf p$ in place of $V\mathbf M$ and $1/\sigma$ in place of $\alpha_G/(|\gamma|M_s)$. The dissipated power, $2\mathcal R_E=|\dot{\mathbf p}_b|^2/(V\sigma)$, is the Joule heating $\int_V (j^2/\sigma)\,d^3r$ of the uniform polarization current $\mathbf j=\dot{\mathbf p}_b/V$ inside the ellipsoid.

\paragraph{Lag torque.} For a static laboratory field, $(d\mathbf E_b/dt)_{\rm body}=-\boldsymbol\Omega_b\times\mathbf E_b$. Because $\Omega\tau_{\chi,i}\ll1$, Eq.~(\ref{eq:Debye_body}) can be solved adiabatically to first order in $\Omega\tau_{\chi,i}$:
\begin{align}
\mathbf p_b&={\boldsymbol \chi} \,\mathbf E_b + \boldsymbol\chi \boldsymbol\tau_{\chi}\,(\boldsymbol\Omega_b\times\mathbf E_b),
\end{align}
where
\begin{align}
(\boldsymbol\chi\boldsymbol\tau_\chi)_{ii}&=\chi_i\tau_{\chi,i}=\frac{\varepsilon_0^2V}{N_i^2\sigma},
\end{align}
is a diagonal matrix; here $\boldsymbol\chi\equiv\operatorname{diag}(\chi_\perp,\chi_\perp,\chi_\parallel)=\varepsilon_0\varepsilon_mV\,\mathbf N_e^{-1}$ and $\boldsymbol\tau_\chi\equiv\operatorname{diag}(\tau_{\chi,\perp},\tau_{\chi,\perp},\tau_{\chi,\parallel})=\tau_\chi\mathbf N_e^{-1}$ are the static polarizability and dipole relaxation-time tensors in the body frame, with $\chi_{\parallel,\perp}=\varepsilon_0V/N_{\parallel,\perp}$ and $\tau_{\chi,\parallel,\perp}=\tau_\chi/N_{\parallel,\perp}$.  The first term gives the conservative torque of Eq.~(\ref{eq:polarization_torque}); the second gives the dissipative (Joule) torque
\begin{equation} \label{eq:N_J}
\mathbf N_{b,J}=\big[\boldsymbol\chi \boldsymbol\tau_\chi \, (\boldsymbol\Omega_b \times \mathbf E_b)\big]\times\mathbf E_b ,
\end{equation}
with $\mathbf N_J=\mathbf R\,\mathbf N_{b,J}$ in the laboratory frame. For an isotropic $\boldsymbol\chi \boldsymbol\tau_\chi$, $\mathbf N_J$ would reduce to $-\chi \tau_\chi E^2 [\boldsymbol\Omega-(\boldsymbol\Omega \cdot \hat{\mathbf E})\hat{\mathbf E}]$, a dissipative torque acting on the component of $\boldsymbol\Omega$ perpendicular to $\mathbf E$.  For the needle the dominant coefficient is $\chi_\parallel \tau_{\chi,\parallel}$, larger than $\chi_\perp \tau_{\chi,\perp}$ by $(N_\perp/N_\parallel)^2\approx38$, so the friction acts mainly on rotations that change the component of $\hat{\mathbf n}$ along $\mathbf E$. One verifies that $-\mathbf N_J\cdot\boldsymbol\Omega=2\mathcal R_E\ge0$. This is the torque responsible for electrorotation of particles in rotating fields \cite{Jones1995}. Unlike the Gilbert torque, which is internal and exchanges angular momentum between $\mathbf S$ and $\mathbf L$, $\mathbf N_J$ transfers angular momentum from the needle to the sources of the external field and dissipates the rotational energy as heat.

\paragraph{Magnitude.} For the cobalt needle whose parameters are discussed in Sec.~\ref{Cobalt-parameters}, [$\sigma\approx1.6\times10^{7}$~S/m where S/m = 1/(ohm m)], we have $\tau_{\chi,\parallel} \approx 7 \times 10^{-18}$~s and $\tau_{\chi,\perp} \approx 1 \times 10^{-18}$~s. Consequently, the ratio of the dissipative to the conservative electric torque is $\Omega \, \tau_{\chi,\parallel} \approx 5 \times 10^{-8}$ at $\Omega/2\pi = 1$~GHz, as noted in Sec.~\ref{subsec:screening}. The friction coefficient evaluates to $\Gamma_J \equiv \chi_\parallel \tau_{\chi,\parallel} E^2 \approx 1 \times 10^{-34}$~N\,m\,s at $E=5\times 10^{7}$~V/m. The corresponding rotational damping rate, $\Gamma_J/I_1 \approx 0.3$~s$^{-1}$, is on the order of $\omega_E^2 \tau_{\chi,\parallel}$, where $\omega_E\approx(p_E E^2/I_1)^{1/2} \approx 2 \times 10^{8}$~s$^{-1}$ is the electrostatic libration frequency. As a result, the angular velocity is damped on a time scale $I_1/\Gamma_J\approx3.3$~s (the rotational energy decays twice as fast), which is considerably slower than the nanosecond Gilbert damping time of the spin. Since $\Gamma_J \propto E^2$, at $E = E_c \approx 8.5 \times 10^{7}$~V/m this rate is larger by a factor of about 2.9.

\paragraph{Magnetic counterpart.} The same conduction electrons dissipate energy through the eddy currents induced by the rotation in the magnetic field \cite{LandauLifshitzElectrodynamics}, with the friction coefficient $\Gamma_{\rm eddy}\sim\sigma B^2Vr^2\approx5\times10^{-32}$~N\,m\,s at $B=0.5$~T (Sec.~\ref{subsec:screening}), i.e., a damping rate $\Gamma_{\rm eddy}/I_1\sim10^{2}$~s$^{-1}$, about $4\times10^{2}$ times larger than the electric one at $E=5\times10^{7}$~V/m. The complete free-electron dissipation function is $\mathcal R_e=(1/2\sigma)\int_Vj^2\,d^3r=\mathcal R_E+\mathcal R_{\rm eddy}$; there is no cross term, because the uniform polarization current is orthogonal to the divergence-free eddy currents, which have no net volume integral. Both channels are negligible on the ns--$\mu$s time scales of Sec.~\ref{sec:Numerical-results}, and they are not included in the numerical results; they can be included as $\mathbf N_J$ in the torque equations of Sec.~\ref{sec:EofMotion} and as a contribution $\Gamma_{\rm eddy}$ to $\boldsymbol\Gamma_{\rm mech}$. They determine, however, the final relaxation of the rotation of an isolated needle, after the spin has relaxed by Gilbert damping, when mechanical friction is absent (e.g., for a needle levitated in vacuum, where GHz rotation of nanoparticles has been achieved \cite{Reimann2018, Ahn2018}).

\subsection{Internal Anisotropy and Conservation of Total Angular Momentum}  \label{subsec:anisotropy-angularmomentum}

A crucial physical consistency requirement of the coupled spin-lattice equations is that internal forces (such as magnetocrystalline and shape anisotropy) cannot alter the total angular momentum of the isolated system, $\mathbf{J}_{\text{tot}} = \mathbf{S} + \mathbf{L}$.

The rate of change of the macro-spin $\mathbf{S}$ due solely to the internal anisotropy field $\mathbf{B}_{\text{ani}} = \frac{2 K_v}{M_s} (\hat{\mathbf{m}} \cdot \hat{\mathbf{n}}) \hat{\mathbf{n}}$ is given by the conservative torque:
\begin{align}
\left( \frac{d\mathbf{S}}{dt} \right)_{\text{ani}} &=
\gamma \mathbf{S} \times \mathbf{B}_{\text{ani}}
\nonumber \\ &=
- \gamma S \, \hat{\mathbf{m}} \times \left( \frac{2 K_v}{M_s} (\hat{\mathbf{m}} \cdot \hat{\mathbf{n}}) \hat{\mathbf{n}} \right) \nonumber \\
&= 2 K_v V (\hat{\mathbf{m}} \cdot \hat{\mathbf{n}}) (\hat{\mathbf{m}} \times \hat{\mathbf{n}}).
\end{align}
Because $H_{\text{ani}}$ depends only on the relative orientation of $\hat{\mathbf m}$ and $\hat{\mathbf n}$, it is invariant under a simultaneous rotation of the spin and the lattice; hence the torque $\mathbf{N}_{\text{ani}}$ exerted by the magnetization on the crystalline lattice must be equal and opposite to the torque experienced by the spin:
\begin{equation}
\mathbf{N}_{\text{ani}} =
-\left( \frac{d\mathbf{S}}{dt} \right)_{\text{ani}} =
- 2 K_v V (\hat{\mathbf{m}} \cdot \hat{\mathbf{n}}) (\hat{\mathbf{m}} \times \hat{\mathbf{n}}).
\end{equation}

In the absence of external fields ($\mathbf{B}=\mathbf{0}, \mathbf{E}=\mathbf{0}$) and damping ($\alpha_G = 0, \boldsymbol{\Gamma}_{\text{mech}} = 0$), the equations of motion for spin and rotational angular momentum are simply $\frac{d\mathbf{S}}{dt} = \left( \frac{d\mathbf{S}}{dt} \right)_{\text{ani}}$ and $\frac{d\mathbf{L}}{dt} = \mathbf{N}_{\text{ani}}$. Summing them yields:
\begin{align}
\frac{d}{dt} \left( \mathbf{S} + \mathbf{L} \right) &= \left( \frac{d\mathbf{S}}{dt} \right)_{\text{ani}} + \mathbf{N}_{\text{ani}} = \mathbf{0}.
\end{align}
Thus, internal anisotropy generates equal and opposite spin and mechanical torques, acting purely as an internal momentum exchange channel between $\mathbf{S}$ and $\mathbf{L}$. Anisotropy makes strictly zero contribution to the rate of change of total angular momentum $\frac{d}{dt}(\mathbf{S} + \mathbf{L})$.\\

\section{Equations of Motion}  \label{sec:EofMotion}

To parameterize 3D rigid-body spatial orientations while ensuring stability across phase space, unit quaternions $\mathbf{q}(t) = (q_0(t), q_1(t), q_2(t), q_3(t))^T$ provide a global, singularity-free formulation of the kinematics \cite{Altmann_05, Evans_77, Omelyan_98}. Below, we detail the complete system of coupled equations of motion that govern spin, rotational angular momentum, and spatial orientation, in both the space-fixed (laboratory) frame and the body-fixed reference frame \cite{Band2018, Chudnovsky2005}.

\subsection{Equations of Motion in the Space-Fixed (Laboratory) Frame}  \label{subsec:EofMotionLab}

In the laboratory frame, all vector quantities ($\mathbf{S}(t), \mathbf{L}(t), \boldsymbol{\Omega}(t), \hat{\mathbf{n}}(t)$) are tracked in Cartesian space coordinates as a function of time \cite{Band2018}. The time evolution of the orientation of the symmetric top is smoothly integrated using the space-fixed unit quaternion $\mathbf{q}(t)$ \cite{Altmann_05, Omelyan_98}.

\begin{enumerate}
    \item \textbf{Space-Fixed Quaternion Kinematics:}
    The 3D orientation matrix $\mathbf{R}(\mathbf{q})$ mapping body coordinates to laboratory coordinates is parameterized by $\mathbf{q}(t)$ \cite{Altmann_05}:
    \begin{equation}  \label{eq:rot_matrix_quaternions}
    \mathbf{R}(\mathbf{q}) = \begin{pmatrix}
    1 - 2(q_2^2 + q_3^2) & 2(q_1 q_2 - q_0 q_3) & 2(q_1 q_3 + q_0 q_2) \\
    2(q_1 q_2 + q_0 q_3) & 1 - 2(q_1^2 + q_3^2) & 2(q_2 q_3 - q_0 q_1) \\
    2(q_1 q_3 - q_0 q_2) & 2(q_2 q_3 + q_0 q_1) & 1 - 2(q_1^2 + q_2^2)
    \end{pmatrix}.
    \end{equation}
    The time derivative of $\mathbf{q}(t)$ driven by the space-fixed rotational angular velocity $\boldsymbol{\Omega}(t) = (\Omega_x, \Omega_y, \Omega_z)^T$ is governed by the linear differential system \cite{Altmann_05, Evans_77, Omelyan_98}:
    \begin{equation}  \label{eq:dq_space_fixed_dt}
    \frac{d\mathbf{q}}{dt} = \frac{1}{2} \mathbf{\Upsilon}(\boldsymbol{\Omega}) \mathbf{q}(t),
    \end{equation}
    where $\mathbf{\Upsilon}(\boldsymbol{\Omega})$ is the space-fixed skew-symmetric rate matrix:
    \begin{equation}  \label{eq:Upsilon}
    \mathbf{\Upsilon}(\boldsymbol{\Omega}) = \begin{pmatrix}
    0 & -\Omega_x & -\Omega_y & -\Omega_z \\
    \Omega_x & 0 & -\Omega_z & \Omega_y \\
    \Omega_y & \Omega_z & 0 & -\Omega_x \\
    \Omega_z & -\Omega_y & \Omega_x & 0
    \end{pmatrix}.
    \end{equation}

    \item \textbf{easy axis Vector and Moment of Inertia Updates:}
    With $\mathbf{q}(t)$ updated in the lab frame, the easy axis vector $\hat{\mathbf{n}}(t) = \mathbf{R}(\mathbf{q}) (0,0,1)^T$ and space-fixed inertia tensor $\boldsymbol{\mathcal{I}}(t)$ are reconstructed dynamically without trigonometric functions or coordinate poles:
    \begin{align}
    \hat{\mathbf{n}}(t) &= \begin{pmatrix} 2(q_1 q_3 + q_0 q_2) \\ 2(q_2 q_3 - q_0 q_1) \\ 1 - 2(q_1^2 + q_2^2) \end{pmatrix}, \\
    \boldsymbol{\mathcal{I}}(t) &= \mathbf{R}(\mathbf{q}) \boldsymbol{\mathcal{I}}_0 \mathbf{R}^T(\mathbf{q}), \\
    \boldsymbol{\Omega}(t) &= \boldsymbol{\mathcal{I}}^{-1}(t) \mathbf{L}(t).
    \end{align}

    \item \textbf{Macro-Spin Dynamics (Space Frame LLG):}
    The time evolution of the macro-spin under precession and   damping in the laboratory frame is given by \cite{LandauLifshitz1935, Gilbert2004, Band2018}:
    \begin{equation}  \label{eq:dSdt.space}
    \frac{d\mathbf{S}}{dt} = \gamma \mathbf{S} \times \mathbf{B}_{\text{eff}} - \frac{\alpha_G}{S} \mathbf{S} \times \left( \frac{d\mathbf{S}}{dt} - \boldsymbol{\Omega} \times \mathbf{S} \right).
    \end{equation}

    \item \textbf{Rotational Angular Momentum Dynamics (Space Frame):}
    The total mechanical torque incorporates external electromagnetic interactions, shape-dependent polarization torque, internal anisotropy recoil, frame-covariant dissipation, and mechanical friction \cite{Chudnovsky2005, Band2018}:
    \begin{equation}  \label{eq:dLdt.space}
    \frac{d\mathbf{L}}{dt} = 
    \mathbf{N}_{\text{elec}} + \mathbf{N}_J + \mathbf{N}_{\text{ani}} + \mathbf{N}_{\text{ME}} + \frac{\alpha_G}{S} \mathbf{S} \times \left( \frac{d\mathbf{S}}{dt} - \boldsymbol{\Omega} \times \mathbf{S} \right) - \boldsymbol{\Gamma}_{\text{mech}} \cdot \boldsymbol{\Omega}.
    \end{equation}
    Here $\boldsymbol{\Gamma}_{\text{mech}}$ is the phenomenological mechanical friction (damping) tensor added to the body's rotational dynamics equation (it may include the eddy-current friction $\Gamma_{\rm eddy}$), $\mathbf N_J$ is the Joule (lag) torque of Eq.~(\ref{eq:N_J}), which is negligible for the parameters of Sec.~\ref{sec:Numerical-results}, the electrostatic torque
    \begin{align}
    \mathbf{N}_{\text{elec}} &= p_E (\mathbf{E} \cdot \hat{\mathbf{n}}) (\hat{\mathbf{n}} \times \mathbf{E}),
    \end{align}
    the anisotropy torque exerted on the lattice is
    \begin{equation} 
    \mathbf{N}_{\text{ani}} = - 2 K_v V (\hat{\mathbf{m}} \cdot \hat{\mathbf{n}}) (\hat{\mathbf{m}} \times \hat{\mathbf{n}}).
    \end{equation}
    and the magnetoelectric torque exerted on the lattice is
    \begin{equation}  \label{eq:N_ME}
    \mathbf{N}_{\text{ME}} = -\gamma\left[\mathbf S\times(\boldsymbol\alpha^{T}\mathbf E)+\mathbf E\times(\boldsymbol\alpha\,\mathbf S)\right],
    \end{equation}
    where $\boldsymbol\alpha(t)=\mathbf R\,\boldsymbol\alpha_b\,\mathbf R^T$ is the magnetoelectric tensor in the laboratory frame ($\boldsymbol\alpha^T=\boldsymbol\alpha$ for the diagonal tensor of Sec.~\ref{subsec:magnetoelectric-tensor}). It follows from $\mathbf N_{\rm ME}=-\partial H_{\rm ME}/\partial\boldsymbol\theta$ for an infinitesimal rotation $\delta\boldsymbol\theta$ of the lattice at fixed $\mathbf S$ and $\mathbf E$, with $H_{\rm ME}=-\gamma\sum_{ij}\alpha_{ij}E_iS_j$ [Eq.~(\ref{eq:HEM})]. The first term is the reaction on the lattice to the precession of $\mathbf S$ in $\mathbf B_{\rm ME}$, and the second arises from the orientation dependence of $\boldsymbol\alpha$. Omitting $\mathbf N_{\rm ME}$ violates exact energy conservation when $\boldsymbol\alpha\neq0$ (the omitted energy change is $\sim10^{-22}$~J for the parameters used here); it is included in our numerical integrations.
\end{enumerate}
Note that the electric torque, which is along $\hat{\mathbf n}\times\hat{\mathbf E}$ with magnitude proportional to $\mathbf E\!\cdot\!\hat{\mathbf n}$, vanishes unless $\mathbf E$ has both a component along and a component perpendicular to the easy axis; likewise the anisotropy torque vanishes unless $\hat{\mathbf m}$ has both components.

\subsection{Equations of Motion in the Body-Fixed Frame}  \label{subsec:EofMotionBody}

In the body-fixed frame aligned with the principal axes of the needle, vector quantities $\mathbf{X}_b$ are defined relative to laboratory vectors $\mathbf{X}_{\text{lab}}$ via the transposed rotation matrix \cite{Altmann_05, Band2018}:
\begin{equation}  \label{eq:body_frame_transform}
\mathbf{X}_b = \mathbf{R}^T(\mathbf{q}) \mathbf{X}_{\text{lab}},
\end{equation}
where $\mathbf{R}(\mathbf{q})$ is given by Eq.~(\ref{eq:rot_matrix_quaternions}). In this reference frame, the principal inertia tensor $\boldsymbol{\mathcal{I}}_0 = \operatorname{diag}(I_1, I_1, I_3)$ and easy axis vector $\hat{\mathbf{n}}_b = (0, 0, 1)^T$ are strictly constant.

\begin{enumerate}
    \item \textbf{Orientation Dynamics (Body-Frame Quaternion Kinematics):}
    Tracking unit quaternion $\mathbf{q}(t)$ in terms of body-fixed angular velocity $\boldsymbol{\Omega}_b = (\Omega_{b1}, \Omega_{b2}, \Omega_{b3})^T$ \cite{Altmann_05, Evans_77, Omelyan_98}:
    \begin{equation}
    \frac{d\mathbf{q}}{dt} = \frac{1}{2} \mathbf{\Xi}(\boldsymbol{\Omega}_b) \mathbf{q}(t),
    \end{equation}
    where $\mathbf{\Xi}(\boldsymbol{\Omega}_b)$ is the skew-symmetric matrix:
    \begin{equation}   \label{eq:Xi}
    \mathbf{\Xi}(\boldsymbol{\Omega}_b) = \begin{pmatrix}
    0 & -\Omega_{b1} & -\Omega_{b2} & -\Omega_{b3} \\
    \Omega_{b1} & 0 & \Omega_{b3} & -\Omega_{b2} \\
    \Omega_{b2} & -\Omega_{b3} & 0 & \Omega_{b1} \\
    \Omega_{b3} & \Omega_{b2} & -\Omega_{b1} & 0
    \end{pmatrix},
    \end{equation}
    and $\boldsymbol{\Omega}_b = \boldsymbol{\mathcal{I}}_0^{-1} \mathbf{L}_b = (L_{b1}/I_1, \, L_{b2}/I_1, \, L_{b3}/I_3)^T$.

    \item \textbf{Transformation of External Fields:}
    The external static fields are mapped into body coordinates at each time instance using Eq.~(\ref{eq:body_frame_transform}):
    \begin{align}
    \mathbf{E}_b(t) &= \mathbf{R}^T(\mathbf{q}(t)) \mathbf{E}_{\text{lab}}, \\
    \mathbf{B}_b(t) &= \mathbf{R}^T(\mathbf{q}(t)) \mathbf{B}_{\text{lab}}.
    \end{align}

    \item \textbf{Macro-Spin Dynamics in Body Frame:}
    The dynamic evolution of magnetization in the rotating reference frame takes the form \cite{LandauLifshitz1935, Gilbert2004, Chudnovsky2005}:
    \begin{equation}
    \left( \frac{d\mathbf{S}_b}{dt} \right)_{\text{body}} = \gamma \mathbf{S}_b \times \mathbf{B}_{\text{eff}, b} + \mathbf{S}_b \times \boldsymbol{\Omega}_b - \frac{\alpha_G}{S} \mathbf{S}_b \times \left( \frac{d\mathbf{S}_b}{dt} \right)_{\text{body}},
    \end{equation}
    where the total body-frame effective magnetic field is:
    \begin{equation}
    \mathbf{B}_{\text{eff}, b} = \mathbf{B}_b + \frac{2 K_v}{M_s} m_{b3} \hat{\mathbf{e}}_z + \mathbf{B}_{\text{ME}, b} + \frac{1}{c^2} (\mathbf{E}_b \times \mathbf{v}_b).
    \end{equation}
    The term $\mathbf S_b\times\boldsymbol\Omega_b=-\boldsymbol\Omega_b\times\mathbf S_b$ arises from $(d\mathbf S/dt)_{\rm body}=d\mathbf S/dt-\boldsymbol\Omega\times\mathbf S$; it is equivalent to the Barnett field $\boldsymbol\Omega/\gamma$. $\mathbf{B}_{\text{ME}, b}$ is defined as $(\mathbf{B}_{\text{ME}, b})_i = \sum_j \alpha_{ji, b} E_{j, b}$ where $E_{j, b}$ is the electric field in the body fixed frame and $\alpha_{ji, b}$ is the linear magnetoelectric tensor in the body fixed frame.

    \item \textbf{Euler Mechanical Equations with Spin Coupling:}
    The rigid-body mechanical dynamics in the body frame obey Euler's equations coupled to the internal and external magnetic and polarization torques \cite{Band2018, Chudnovsky2005}:
    \begin{align}
    &\boldsymbol{\mathcal{I}}_0 \left(\frac{d\boldsymbol{\Omega}_b}{dt}\right)_{\text{body}} + \boldsymbol{\Omega}_b \times (\boldsymbol{\mathcal{I}}_0 \boldsymbol{\Omega}_b) = \mathbf{N}_{b, \text{elec}} + \mathbf N_{b,J} \nonumber \\
     &\quad + \mathbf{N}_{b, \text{ani}} + \mathbf{N}_{b,\text{ME}} + \frac{\alpha_G}{S}\mathbf S_b \times \left( \frac{d\mathbf{S}_b}{dt} \right)_{\text{body}} - \boldsymbol{\Gamma}_{b} \cdot \boldsymbol{\Omega}_b,
    \end{align}
    with electric torque and internal anisotropy torque given by:
    \begin{align}
    \mathbf{N}_{b, \text{elec}} &= p_E E_{b3} (\hat{\mathbf{e}}_z \times \mathbf{E}_b) ,\\
    \mathbf{N}_{b, \text{ani}} &= - 2 K_v V m_{b3} (m_{b2}, -m_{b1}, 0)^T,
    \end{align}
    $\boldsymbol\Gamma_b=\mathbf R^T\boldsymbol\Gamma_{\rm mech}\mathbf R$, $\mathbf N_{b,J}$ is given by Eq.~(\ref{eq:N_J}), and the magnetoelectric torque, $\mathbf N_{b,\rm ME}=\mathbf R^T\mathbf N_{\rm ME}$ [Eq.~(\ref{eq:N_ME})], is
    \begin{equation}
    \mathbf N_{b,\text{ME}}=-\gamma\left[\mathbf S_b\times(\boldsymbol\alpha_b^{T}\mathbf E_b)+\mathbf E_b\times(\boldsymbol\alpha_b\,\mathbf S_b)\right],
    \end{equation}
    with the constant body-frame tensor $\boldsymbol\alpha_b$.
\end{enumerate}

\section{Dimensionless Equations of Motion}  \label{EofMDimensionless}

To avoid propagation of large unscaled dimensional variables during numerical integration [e.g., $|\mathbf{S}| \approx 4.15 \times10^{-28}\text{ J}\cdot\text{s}$, $|\mathbf{L}| \approx 4.06 \times 10^{-32}\text{ J}\cdot\text{s}$, or $|\boldsymbol{\Omega}| \sim 10^2\text{ s}^{-1}$ (these are initial values; during the dynamics of Sec.~\ref{sec:Numerical-results} $|\frac{\mathbf{L}}{S}|$ reaches $\sim10^{2}$--$10^{3}$ and $|\boldsymbol\Omega| \sim 10^{8}$--$10^{9}\text{ s}^{-1}$], we recast the complete coupled system into dimensionless form \cite{Band2018}. 

We define the normalized spin vector $\hat{\mathbf{s}}$, dimensionless rotational angular momentum vector $\hat{\mathbf{l}}$, and dimensionless time variable $\tau$:
\begin{equation}
\hat{\mathbf{s}} = \frac{\mathbf{S}}{S}, \quad \hat{\mathbf{l}} = \frac{\mathbf{L}}{S}, \quad \tau = \omega_0 t,
\end{equation}
where $S = |\mathbf{S}| = M_s V / |\gamma|$ is the conserved spin magnitude ($|\hat{\mathbf{s}}| = 1$), and $\omega_0$ is a characteristic frequency scale chosen as $\omega_0 = |\gamma| B_0$ (for a reference magnetic field magnitude $B_0$) or as the natural anisotropy precession frequency $\omega_{\text{ani}} = \frac{2 K_v V}{S} = \frac{2 |\gamma| K_v}{M_s}$ ($\approx 2.6 \times 10^{11}$~s$^{-1}$ for Co, with $g\simeq2.2$).

\subsection{Dimensionless Scaling Parameters}  \label{Dimensionless-parameters}

We define the following dimensionless parameters:
\begin{align}
\tilde{\mathbf{B}} &= \frac{\mathbf{B}}{B_0}, \quad \tilde{\mathbf{E}} = \frac{\mathbf{E}}{E_0}, \\
\tilde{\boldsymbol{\mathcal{I}}} &= \frac{\omega_0}{S} \boldsymbol{\mathcal{I}} \quad \Longrightarrow \quad \tilde{\boldsymbol{\mathcal{I}}}_0 = \frac{\omega_0}{S} \operatorname{diag}(I_1, I_1, I_3), \\
\tilde{\boldsymbol{\Omega}} &= \frac{\boldsymbol{\Omega}}{\omega_0} = \tilde{\boldsymbol{\mathcal{I}}}^{-1} \hat{\mathbf{l}}, \\
\kappa &= \frac{2 K_v V}{S \omega_0} = \frac{2 K_v}{M_s B_0}, \\
\tilde p_E &= \frac{p_E E_0^2}{S \omega_0}, \\
(\boldsymbol{\beta}_{\text{ME}})_j &= \frac{|\gamma|}{\omega_0} \sum_i \alpha_{ij} E_i, \quad \tilde{\boldsymbol\alpha} = \frac{|\gamma|E_0}{\omega_0}\boldsymbol\alpha \;\Longrightarrow\; \boldsymbol\beta_{\rm ME}=\tilde{\boldsymbol\alpha}^{T}\tilde{\mathbf E}, \\
\eta_{\text{SO}} &= \frac{v_0 E_0}{c^2 B_0}, \\
\tilde{\boldsymbol{\Gamma}}_{\text{mech}} &= \frac{\boldsymbol{\Gamma}_{\text{mech}}}{S},
\end{align}
where $v_0$ is a characteristic velocity scale for normalization ($\tilde{\mathbf{v}} = \mathbf{v}/v_0$). $B_0$ and $E_0$ are reference field amplitudes (not the static fields $\mathbf B_0$, $\mathbf E_0$ of Secs.~\ref{sec:depolarization_tensor} and \ref{subsec:magnetoelectric-tensor}). The Joule torque enters the dimensionless equations as $\tilde{\mathbf N}_J=\mathbf N_J/(S\omega_0)$ and is omitted there because it is negligible.
In Sec.~\ref{sec:Numerical-results} we take $\omega_0=|\gamma|B_0$, with $B_0$ the applied magnetic field of each run, and $E_0=5\times10^{7}$~V/m, the applied electric field. For the runs at $B_0=0.5$~T, $\omega_0\approx9.67\times10^{10}$~s$^{-1}$, $\kappa=2K_v/(M_sB_0)\approx2.65$, $\tilde I_1=\omega_0I_1/S\approx9.0\times10^{4}$, $\tilde I_3\approx1.06\times10^{4}$ and $\tilde\Gamma_{\rm mech}/\tilde I_1\approx10^{-46}$. For the low-field runs, $B_0=5~\mu$T, $\omega_0\approx9.67\times10^{5}$~s$^{-1}$ and $\kappa\approx2.65\times10^{5}$. Since $S\omega_0=M_sVB_0=\mu B_0$, the electric parameter is $\tilde p_E=p_EE_0^2/(\mu B_0)=(E_0/E_c)^2$, i.e., $\approx0.35$ at $0.5$~T and $\approx3.5\times10^{4}$ at $5~\mu$T. The velocity scale $v_0$ enters only through the product $\eta_{\rm SO}\tilde{\mathbf v}=E_0\mathbf v/(c^2B_0)$, so its value is immaterial; for $v_0=1$~m/s, $\eta_{\rm SO}=E_0v_0/(c^2B_0)\approx1.1\times10^{-9}$ at $0.5$~T, and the motional field vanishes identically when $\mathbf r_{\rm eff}=\mathbf 0$ and $\mathbf v_{\rm cm}=\mathbf 0$ (Sec.~\ref{sec:Hamiltonian}).

\subsection{Space-Fixed Dimensionless Coupled Equations}

In terms of dimensionless time $\tau$, space-fixed normalized spin $\hat{\mathbf{s}}$, normalized rotational angular momentum $\hat{\mathbf{l}}$, easy axis vector $\hat{\mathbf{n}} = \mathbf{R}(\mathbf{q})\hat{\mathbf{e}}_z$, and unit quaternion $\mathbf{q}$, the explicit coupled system becomes:

\begin{enumerate}
    \item \textbf{Quaternion Kinematics}
    \begin{equation}
    \frac{d\mathbf{q}}{d\tau} = \frac{1}{2} \mathbf{\Upsilon}(\tilde{\boldsymbol{\Omega}}) \mathbf{q}(\tau),
    \end{equation}
    with $\tilde{\boldsymbol{\Omega}}(\tau) = \tilde{\boldsymbol{\mathcal{I}}}^{-1}(\tau) \hat{\mathbf{l}}(\tau)$ and $\tilde{\boldsymbol{\mathcal{I}}}(\tau) = \mathbf{R}(\mathbf{q}) \tilde{\boldsymbol{\mathcal{I}}}_0 \mathbf{R}^T(\mathbf{q})$.

    \item \textbf{Dimensionless Effective Magnetic Field}
    \begin{equation}
    \tilde{\mathbf{B}}_{\text{eff}} = \tilde{\mathbf{B}} - \kappa (\hat{\mathbf{s}} \cdot \hat{\mathbf{n}}) \hat{\mathbf{n}} + \boldsymbol{\beta}_{\text{ME}} + \eta_{\text{SO}} (\tilde{\mathbf{E}} \times \tilde{\mathbf{v}}).
    \end{equation}

    \item \textbf{Dimensionless Macro-Spin LLG Dynamics:}
    Expanding the implicit damping term yields the explicit dimensionless macro-spin LLG equation of motion:
    \begin{align}  \label{eq:dSdt.space.dimensionless}
    (1 + \alpha_G^2) \frac{d\hat{\mathbf{s}}}{d\tau} &= -\hat{\mathbf{s}} \times \tilde{\mathbf{B}}_{\text{eff}} + \alpha_G \hat{\mathbf{s}} \times (\hat{\mathbf{s}} \times \tilde{\mathbf{B}}_{\text{eff}}) \nonumber \\
    & - \alpha_G \hat{\mathbf{s}} \times (\hat{\mathbf{s}} \times \tilde{\boldsymbol{\Omega}}) - \alpha_G^2\, \hat{\mathbf{s}} \times \tilde{\boldsymbol{\Omega}}.
    \end{align}
    Note that (a) the first term on the right-hand side of Eq.~(\ref{eq:dSdt.space.dimensionless}) has a minus sign because $\gamma$ is negative, while the damping terms that are proportional to $\alpha_G$ carry the opposite sign (relaxation toward $-\tilde{\mathbf B}_{\rm eff}$, i.e. $\boldsymbol\mu\parallel\mathbf B_{\rm eff}$, because $\hat{\mathbf m}=-\hat{\mathbf s}$; the expansion follows by crossing the implicit equation with $\hat{\mathbf s}$ and using $\hat{\mathbf s}\times(\hat{\mathbf s}\times\mathbf u)=-\mathbf u$, for $\mathbf u\perp\hat{\mathbf s}$), and (b) the Gilbert term generates $\tilde{\boldsymbol\Omega}$-dependent terms.  These $\tilde{\boldsymbol\Omega}$-dependent terms serve to rotate the spin a bit from the direction of $-\tilde{\mathbf{B}}_{\text{eff}}$, but these terms, in practice, are very small since $\alpha_G \ll 1$.

    \item \textbf{Dimensionless Rotational Angular Momentum Dynamics}
    Using Eq.~(\ref{eq:dLdt.space}) and changing to dimensionless variables we obtain the dimensionless rotational angular momentum equation of motion:
    \begin{equation}  \label{eq:dLdt.space.dimensionless}
    \frac{d\hat{\mathbf{l}}}{d\tau} =
    \tilde{\mathbf{N}}_{\text{elec}} + \tilde{\mathbf{N}}_{\text{ani}} + \tilde{\mathbf{N}}_{\text{ME}} + \alpha_G \hat{\mathbf{s}} \times \left( \frac{d\hat{\mathbf{s}}}{d\tau} - \tilde{\boldsymbol{\Omega}} \times \hat{\mathbf{s}} \right) - \tilde{\boldsymbol{\Gamma}}_{\text{mech}} \cdot \tilde{\boldsymbol{\Omega}},
    \end{equation}
    where the dimensionless electric and internal anisotropy torques are:
    \begin{align}
      \tilde{\mathbf{N}}_{\text{elec}} &= \tilde p_E (\tilde{\mathbf{E}} \cdot \hat{\mathbf{n}}) (\hat{\mathbf{n}} \times \tilde{\mathbf{E}}), \\
    \tilde{\mathbf{N}}_{\text{ani}} &= - \kappa (\hat{\mathbf{s}} \cdot \hat{\mathbf{n}}) (\hat{\mathbf{s}} \times \hat{\mathbf{n}}), \\
    \tilde{\mathbf{N}}_{\text{ME}} &= \hat{\mathbf s}\times\boldsymbol\beta_{\rm ME} + \tilde{\mathbf E}\times(\tilde{\boldsymbol\alpha}\,\hat{\mathbf s}),
    \end{align}
\end{enumerate}

\subsection{Body-Fixed Dimensionless Coupled Equations}

Transforming to the rotating reference frame aligned with the principal axes of the symmetric-top needle ($\hat{\mathbf{n}}_b = (0,0,1)^T$):

\begin{enumerate}
    \item \textbf{Body-Frame Quaternion Kinematics}
    \begin{equation}
    \frac{d\mathbf{q}}{d\tau} = \frac{1}{2} \mathbf{\Xi}(\tilde{\boldsymbol{\Omega}}_b) \mathbf{q}(\tau),
    \end{equation}
    where $\tilde{\boldsymbol{\Omega}}_b = \tilde{\boldsymbol{\mathcal{I}}}_0^{-1} \hat{\mathbf{l}}_b = (\hat{l}_{b1}/\tilde{I}_{1}, \, \hat{l}_{b2}/\tilde{I}_{1}, \, \hat{l}_{b3}/\tilde{I}_{3})^T$.

    \item \textbf{Transformation of Dimensionless Fields}
    Laboratory field inputs map directly via the inverse rotation matrix:
    \begin{align}
    \tilde{\mathbf{E}}_b(\tau) &= \mathbf{R}^T(\mathbf{q}(\tau)) \tilde{\mathbf{E}}_{\text{lab}}, \\
    \tilde{\mathbf{B}}_b(\tau) &= \mathbf{R}^T(\mathbf{q}(\tau)) \tilde{\mathbf{B}}_{\text{lab}}.
    \end{align}

    \item \textbf{Dimensionless Body-Frame Macro-Spin Dynamics}
    \begin{equation} \label{eq:dsb.dimless}
    (1 + \alpha_G^2) \left( \frac{d\hat{\mathbf{s}}_b}{d\tau} \right)_{\text{body}} = -\hat{\mathbf{s}}_b \times \tilde{\mathbf{F}}_b + \alpha_G \hat{\mathbf{s}}_b \times (\hat{\mathbf{s}}_b \times \tilde{\mathbf{F}}_b),
    \end{equation}
    with $\tilde{\mathbf{F}}_b\equiv\tilde{\mathbf{B}}_{\text{eff}, b} -\tilde{\boldsymbol\Omega}_b$. This follows from the dimensional body-frame spin equation of Sec.~\ref{subsec:EofMotionBody}; the term $-\tilde{\boldsymbol\Omega}_b$ in $\tilde{\mathbf F}_b$ is the Barnett field.  Here the effective magnetic field in the body-fixed frame is:
    \begin{equation}
    \tilde{\mathbf{B}}_{\text{eff}, b} = \tilde{\mathbf{B}}_b - \kappa s_{b3} \hat{\mathbf{e}}_z + \boldsymbol{\beta}_{\text{ME}, b} + \eta_{\text{SO}}(\tilde{\mathbf{E}}_b \times \tilde{\mathbf{v}}_b).
    \end{equation}

    \item \textbf{Dimensionless Euler Equations in Body Frame}
    \begin{align}
    \tilde{\boldsymbol{\mathcal{I}}}_0 \left(\frac{d\tilde{\boldsymbol{\Omega}}_b}{d\tau}\right)_{\text{body}} + \tilde{\boldsymbol{\Omega}}_b \times (\tilde{\boldsymbol{\mathcal{I}}}_0 \tilde{\boldsymbol{\Omega}}_b) = 
    \tilde{\mathbf{N}}_{b, \text{elec}} + \tilde{\mathbf{N}}_{b, \text{ani}} + \tilde{\mathbf{N}}_{b, \text{ME}} \nonumber \\
     + \alpha_G\hat{\mathbf s}_b\times\left( \frac{d\hat{\mathbf{s}}_b}{d\tau} \right)_{\text{body}} - \tilde{\boldsymbol{\Gamma}}_b \cdot \tilde{\boldsymbol{\Omega}}_b,
    \end{align}
    with electric torque $\tilde{\mathbf{N}}_{b, \text{elec}} = \tilde p_E \tilde{E}_{b3} (\hat{\mathbf{e}}_z \times \tilde{\mathbf{E}}_b)$, internal anisotropy torque $\tilde{\mathbf{N}}_{b, \text{ani}} = -\kappa s_{b3} (s_{b2}, -s_{b1}, 0)^T$, and magnetoelectric torque $\tilde{\mathbf{N}}_{b, \text{ME}} = \hat{\mathbf s}_b\times\boldsymbol\beta_{\text{ME}, b} + \tilde{\mathbf E}_b\times(\tilde{\boldsymbol\alpha}_b\,\hat{\mathbf s}_b)$.
\end{enumerate}

\section{Single-Domain Cobalt Needle Parameters}  \label{Cobalt-parameters}

To model a physical magnetic gyroscope, we consider a single-domain hexagonal close-packed (hcp) cobalt needle fashioned as a prolate spheroid (symmetric-top) with major semi-axis length $a$ along the $z$-axis (easy axis $\hat{\mathbf{n}}$) and transverse semi-axes $b = c = r$.

For hcp cobalt at room temperature ($T = 300\text{ K}$):
\begin{itemize}
    \item \textbf{Mass density:} $\rho \approx 8.90 \times 10^3 \text{ kg/m}^3$ \cite{Lide2005}.
    \item \textbf{Saturation magnetization:} $M_s \approx 1.42 \times 10^6 \text{ A/m}$ ($\mu_0 M_s \approx 1.79 \text{ T}$) \cite{Cullity2008, Bozorth1993}.
    \item \textbf{Magnetocrystalline anisotropy constant:} Uniaxial $K_{u1} \approx 4.5 \times 10^5 \text{ J/m}^3$ along the $c$-axis $[0001]$ \cite{Sucksmith1954, Chikazumi1997}.
    \item \textbf{Gilbert damping constant:} $\alpha_G \approx 0.005 - 0.01$ \cite{Oogane2006, Woltersdorf2009}.
    \item \textbf{Gyromagnetic ratio:} $\gamma \approx -1.76 \times 10^{11} \text{ rad/(s}\cdot\text{T)}$ for $g \approx 2.0$ (free electron value without QED radiative corrections).  For Co \cite{Getzlaff2007}, $g\simeq2.2$, i.e., $\gamma=-g\mu_B/\hbar\approx -1.9347 \times 10^{11} \text{ rad/(s}\cdot\text{T)}$; this value is used throughout the paper (text, figures, and numerical code).
    \item \textbf{Static (DC) Depolarization Tensor and Polarizability ($\mathbf{N}_e, \mathbf{N}_e^{-1}$):} In static electric fields, conduction electrons redistribute to screen the internal field of metallic cobalt completely ($\mathbf{E}_{\text{in}} \to \mathbf{0}$) \cite{Jackson1999}. The polarization response and induced electrostatic torque $\mathbf{N}_{\text{elec}} = \mathbf{p} \times \mathbf{E}$ are strictly governed by the shape-dependent depolarizing factors $N_x, N_y, N_z$ of the prolate spheroid \cite{Stratton1941, LandauLifshitzElectrodynamics}:
    \begin{equation}
    \chi_{e, ii} \equiv \frac{1}{N_i}, \quad (i = x, y, z),
    \end{equation}
    where for eccentricity $\xi = \sqrt{1 - (r/a)^2}$:
    \begin{align}
    N_z &= \frac{1 - \xi^2}{2 \xi^3} \left( \ln\left( \frac{1+\xi}{1-\xi} \right) - 2\xi \right), \\
    N_x &= N_y = \frac{1 - N_z}{2}.
    \end{align}
    For a cobalt needle geometry ($a/r = 4$, $\xi \approx 0.9682$), the DC depolarizing factors are $N_z = N_\parallel \approx 0.0754$ and $N_x = N_y = N_\perp \approx 0.4623$. This yields dimensionless effective polarizability components $\chi_{e,zz} \approx 13.26$ and $\chi_{e,xx}=\chi_{e,yy} \approx 2.16$, providing a shape polarizability parameter
    \begin{align}
    p_E &= \varepsilon_0 \varepsilon_m V
    \Big( \frac{1}{N_\parallel} - \frac{1}{N_\perp} \Big) \approx
    11.1 \, \varepsilon_0 \varepsilon_m V \nonumber \\
    &\approx 5.6\times10^{-33}\ \text{C\,m}^2/\text{V}.
    \end{align}
    With $N_z=0.0754$, the shape anisotropy constant is $K_{\rm shape}=\tfrac14\mu_0M_s^2(1-3N_z)\approx4.9\times10^{5}$~J/m$^3$, so $K_v\approx9.4\times10^{5}$~J/m$^3$ and the anisotropy field is $2K_v/M_s\approx1.32$~T.
\end{itemize}

\subsection{Linear Magnetoelectric Tensor ($\alpha_{ij}$) for Prolate Cobalt Needle}  \label{subsec:magnetoelectric-tensor}

In bulk single-crystal hcp cobalt ($D_{6h}$ point group), global spatial inversion symmetry prohibits intrinsic linear magnetoelectric (ME) coupling ($\alpha_{ij} = 0$) in zero external bias electric field \cite{Fiebig2005, Rivera1994}. However, for a nanoscale prolate cobalt needle, surface spin-dependent screening can generate an effective linear magnetoelectric tensor $\boldsymbol{\alpha}$. An applied DC electric field induces surface charge accumulation over a screening length $\lambda_F \approx 0.5 \text{ \AA}$ (Thomas--Fermi length in Co) \cite{Duan2008, Zhang2008}. This surface field modifies the local density of states and spin-orbit coupling, producing a surface magnetoelectric tensor $\boldsymbol{\alpha}^{\text{surf}}$ \cite{Zhang2008, Weisheit2007}:
    \begin{equation}  \label{eq:magnetoelectric_alpha}
    \alpha_{ii} \sim \frac{\lambda_F}{a_i} \alpha_{ii}^{\text{surf}},
    \end{equation}
    where $a_i$ is the semi-axis along the field component $E_i$ ($a_z=a$, $a_x=a_y=r$): a field along axis $i$ accumulates screening charge on the faces perpendicular to that axis, and the screening layer of thickness $\lambda_F$ on these faces occupies a fraction $\sim\lambda_F/a_i$ of the volume. [If $\boldsymbol\alpha^{\rm surf}$ is defined per unit area, $\mu_0\Delta M_{\rm area}=\alpha_sE$ (units of s), as in Ref.~\cite{Duan2008}, then $\lambda_F$ cancels and $\alpha_{ii}\sim\alpha_s/(2a_i)$; the local enhancement $1/N_i$ of the surface field, Sec.~\ref{subsec:screening}, multiplies either estimate.]
    Due to the $D_{\infty h}$ symmetry of the prolate symmetric-top needle aligned along $z$, the tensor $\boldsymbol{\alpha}$ is diagonal with two independent non-zero components:
    \begin{equation}
    \boldsymbol{\alpha} = \begin{pmatrix}
    \alpha_{\perp} & 0 & 0 \\
    0 & \alpha_{\perp} & 0 \\
    0 & 0 & \alpha_{\parallel}
    \end{pmatrix}.
    \end{equation}
    Note that the magnetoelectric interaction is very small because of the Thomas--Fermi screening, and it plays almost no role in the dynamics that will be presented in the next section. Moreover, a linear magnetoelectric effect requires broken space-inversion as well as broken time-reversal symmetry. The magnetization breaks time reversal, but a free prolate spheroid of centrosymmetric hcp Co is inversion symmetric: in a uniform $\mathbf E$ one tip accumulates and the other loses electrons, so the surface contributions of the two ends are equal and opposite and the net linear coefficient vanishes. The values given below should therefore be regarded as upper bounds, relevant only when inversion symmetry is broken (e.g., a needle on a substrate, or one end oxidized or capped).  For the parameters given below the effect is negligible even when inversion symmetry is broken.

A bulk, field-induced contribution is not included. With $\alpha_{ij}=\mu_0\,\partial M_i/\partial E_j$ (units s/m), such a term would require a contribution to the anisotropy that is odd in $\mathbf E$; in centrosymmetric hcp Co the anisotropy constant is even in $\mathbf E$, so $\partial K_{u1}/\partial E=0$ at $E=0$, and the leading bulk effects are quadratic in $\mathbf E$ and do not give a linear tensor $\boldsymbol\alpha$.

\textbf{Final Numerical Value for $\boldsymbol{\alpha}$ Tensor:}  
Using surface screening density functional calculations for Co(0001) interfaces ($\alpha_{zz}^{\text{surf}} \approx 1.2 \times 10^{-11} \text{ s/m}$) \cite{Zhang2008, Weisheit2007} scaled by the factors $\lambda_F / a \approx (0.05 \text{ nm}) / (60 \text{ nm}) \approx 8.3 \times 10^{-4}$ and $\lambda_F/r\approx3.3\times10^{-3}$, and assuming $\alpha^{\rm surf}_{xx}\approx\alpha^{\rm surf}_{zz}$:
\begin{align}
\alpha_{\parallel} &= \alpha_{zz} \approx 1.0 \times 10^{-14} \text{ s/m} \nonumber \\
& (\approx 3.0 \times 10^{-6} \text{ Gaussian / SI}), \\
\alpha_{\perp} &= \alpha_{xx} = \alpha_{yy} \approx 4.0 \times 10^{-14} \text{ s/m}.
\end{align}
Thus, the explicit body-fixed linear magnetoelectric tensor for the symmetric-top cobalt needle is:
\begin{equation}
\boldsymbol{\alpha}_b = \begin{pmatrix}
4.0\times10^{-14} & 0 & 0 \\
0 & 4.0\times10^{-14} & 0 \\
0 & 0 & 1.0 \times 10^{-14}
\end{pmatrix} \text{ s/m}.
\end{equation}

For a prolate needle geometry of radius $r = 15 \text{ nm} (= b = c)$ and length $L = 2a = 120 \text{ nm}$ (aspect ratio $a/r = 4$, keeping the radius safely below the single-domain limit $R_{c} \approx 20\text{--}30\text{ nm}$ \cite{Kittel1946}):
\begin{align}
V &= \frac{4}{3}\pi a r^2 \approx 5.65 \times 10^{-23} \text{ m}^3, \\
m &= \rho V \approx 5.03 \times 10^{-19} \text{ kg}, \\
I_1 = I_2 &= \frac{1}{5} m (a^2 + r^2) \approx 3.85 \times 10^{-34} \text{ kg}\cdot\text{m}^2, \\
I_3 &= \frac{2}{5} m r^2 \approx 4.53 \times 10^{-35} \text{ kg}\cdot\text{m}^2.
\end{align}
Note that $I_1 = I_2 \neq I_3$, forming a prolate symmetric top.  Total spin magnitude is $S = \frac{M_s V}{|\gamma|} \approx 4.15 \times 10^{-28} \text{ J}\cdot\text{s} \approx 3.9 \times 10^{6} \hbar$ (for $|\gamma|=1.9347\times10^{11}$ rad/(s$\cdot$T), $g=2.2$).  The magnetic moment is $\mu = M_s V \approx 8.0 \times 10^{-17}$~J/T.

\section{Numerical Results}  \label{sec:Numerical-results}

This section presents calculations of the dynamics of a cobalt symmetric-top magnetic needle (gyroscope) with semi-major axis $a = 60$ nm and semi-minor axis $r = 15$ nm, in the presence of a static magnetic field.  The case of a static magnetic field and electric field is also considered.  We use a value $\alpha_G = 0.01$ for the LLG parameter, and a very small mechanical damping rate $\Gamma_{\rm mech}/I_1 = 1.0 \times 10^{-35}$~s$^{-1}$ [in the dimensionless equations this corresponds to $\tilde\Gamma_{\rm mech}/\tilde I_1=\Gamma_{\rm mech}/(\omega_0I_1)\approx10^{-46}$], hence it is entirely negligible in our calculations; it is also much smaller than the unavoidable free-electron damping rates of Sec.~\ref{subsec:Joule}, which are themselves negligible on the time scales considered here. The initial conditions in the space-fixed frame are as follows: the initial normalized spin is $\mathbf{s}(0) = (0.2, -0.3064, 0.9307)$, the initial normalized rotational angular momentum is $\mathbf{l}(0) = (8.439,-2.482,4.299) \times 10^{-5}$, and the initial direction of the easy axis is $\hat{\mathbf{n}}(0) = (0, -\frac{1}{2}, \frac{\sqrt{3}}{2})$, i.e., the initial tilt angle is $\theta = \pi/6$.
Note that the initial spin is not aligned with the direction of the initial effective magnetic field $\mathbf{B}_{\text{eff}}(0)$ given in Eq.~(\ref{eq:B_eff}).  The gyro is released at $t = 0$ with the finite initial rotational angular momentum and the dynamics begins.

\begin{figure}
\includegraphics[width=0.5\textwidth]{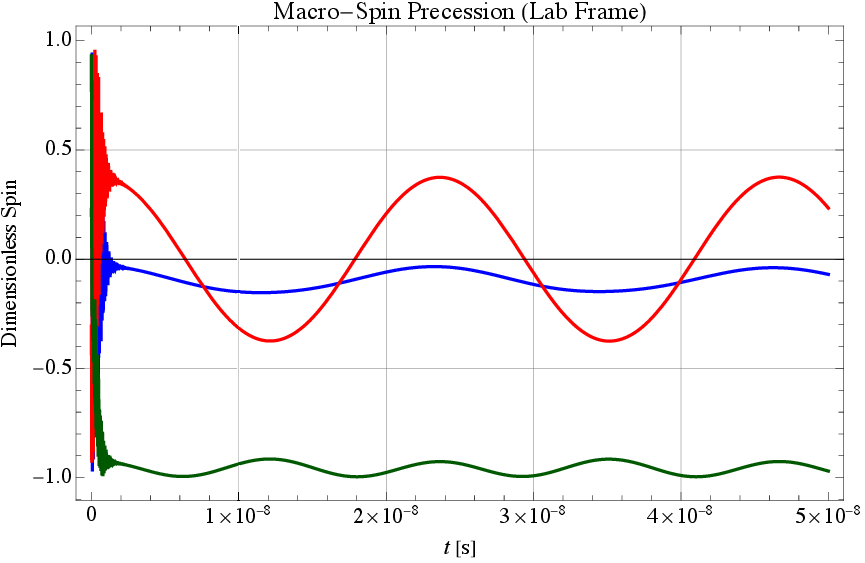}
\caption{\footnotesize  Normalized spin, ${\mathbf s}(t)$, versus time with $B = 0.5$ T, ${\hat {\mathbf B}} = (0.0995, 0., 0.995)$, no electric field, initial tilt quaternion $\mathbf{q}(0) = (\cos(\pi/12), \sin(\pi/12), 0, 0)^T$ corresponding to a rotation by $\pi/6$ about the $x$ axis, i.e., the $z$-$y$-$z$ Euler angles $\theta=\pi/6$, $\phi=-\pi/2$, $\psi=\pi/2$, laboratory frame ${\mathbf s}(0) = (0.2,-0.3064,0.9307)$, ${\mathbf l}(0) = (8.4387, -2.482, 4.299) \times 10^{-5}$, and easy axis ${\mathbf n}(0) = (0,-0.5, 0.8660)$, the final time is taken to be $t_f = 50$ ns, and the Gilbert damping parameter is taken to be $\alpha_G = 0.01$.  The blue curve is $s_x(t)$, the red curve is  $s_y(t)$, and the green curve is  $s_z(t)$.}
\label{Fig_magnetic_needle_spin}
\end{figure}

\begin{figure}
\includegraphics[width=0.45\textwidth]{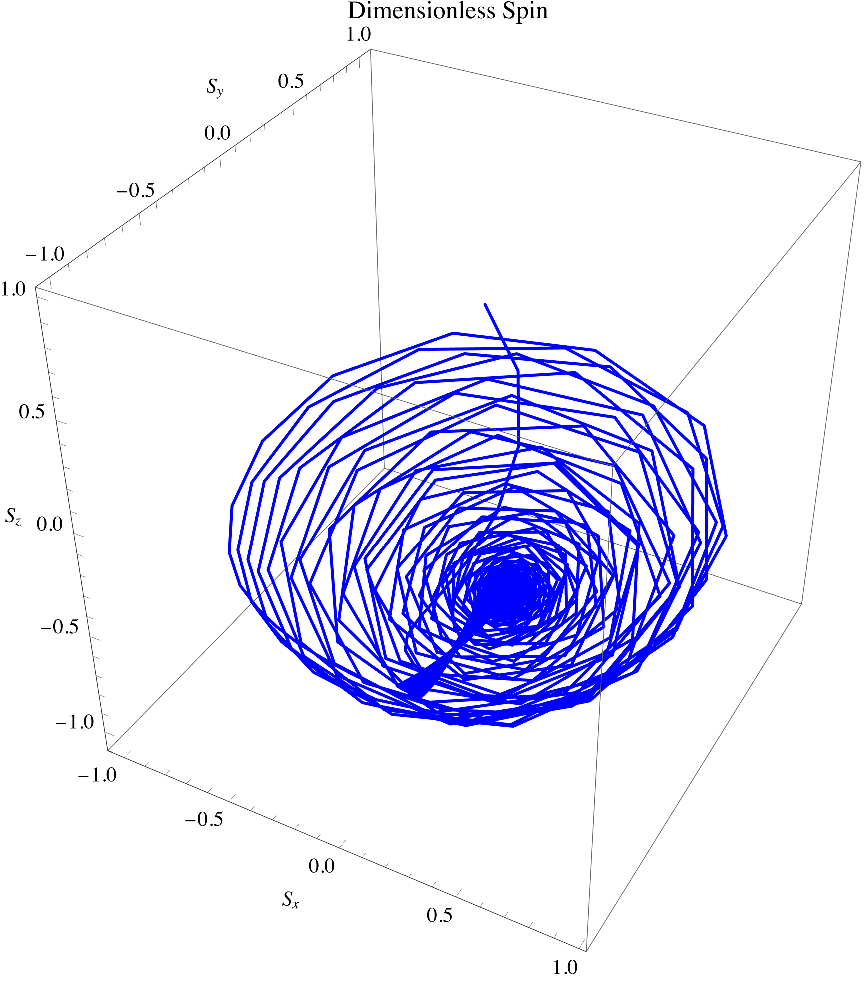}
\caption{\footnotesize  Parametric plot of the normalized spin, ${\mathbf s}(t)$.  The integration time used is $t_f = 0.5 \, \mu$s.  The initial conditions and the parameters used in the calculation are the same as in Fig.~\ref{Fig_magnetic_needle_spin}.}
\label{Fig.needle_spin}
\end{figure}

Figure~\ref{Fig_magnetic_needle_spin} shows the normalized spin of a magnetic gyro, ${\mathbf s}(t)$, versus time in a static magnetic field with $B = 0.5$ T and ${\hat {\mathbf B}} = (0.0995, 0.0, 0.995)$, and no electric field, for a given set of initial conditions as specified in the caption. (See also Fig.~\ref{Fig.needle_spin}.) Near $t = 0$, the LLG dissipation quickly changes ${\mathbf s}(t)$.  Only at very small times, $t < \tau_G$, does the LLG dissipation dynamics cause severe oscillation of the spin vector. Throughout the dynamics, $s_x^2(t) + s_y^2(t) + s_z^2(t) = 1$.  The components $s_x(t)$ and $s_y(t)$ oscillate with a period of about 20 nanoseconds, however the period of oscillations of $s_z(t)$ is about half this period.  This type of motion is called precession accompanied by nutation.

\begin{figure}
\includegraphics[width=0.5\textwidth]{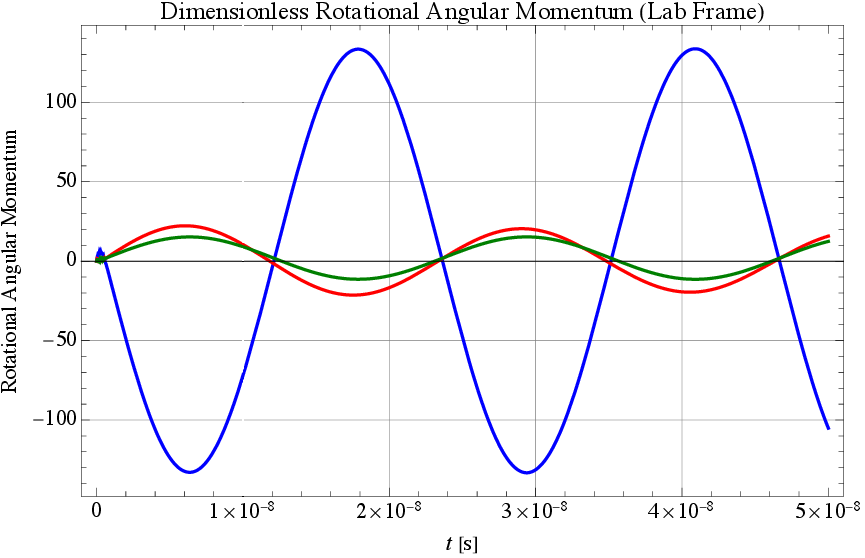}
\caption{\footnotesize  Normalized rotational angular momentum, ${\mathbf l}(t)$, versus time in a magnetic field.   The blue curve is $l_x(t)$, the red curve is  $l_y(t)$, and the green curve is $l_z(t)$.  The initial conditions and the parameters used in the calculation are the same as in Fig.~\ref{Fig_magnetic_needle_spin}.}
\label{Fig_magnetic_needle_rotational_ang_momentum}
\end{figure}

Figure~\ref{Fig_magnetic_needle_rotational_ang_momentum} plots the normalized rotational angular momentum ${\mathbf l}(t)$ of the magnetic gyro versus time.
The components of ${\mathbf l}(t)$ oscillate with the same period of approximately 20 nanoseconds, which is the same as the periods of $s_x(t)$ and $s_y(t)$.
We can understand the large peak rotational angular momentum obtained in Fig.~\ref{Fig_magnetic_needle_rotational_ang_momentum} as follows.  The theoretical maximum kinetic energy occurs when the easy axis and magnetization align with the external field ($\hat{\mathbf{n}} = \hat{\mathbf{m}} = \hat{\mathbf{z}}$), where the Zeeman energy is $\mathcal E_{Z} = -\boldsymbol{\mu} \cdot \mathbf{B} = - M_s V B$, while simultaneously minimizing the internal anisotropy potential energy, $\mathcal E_{K_v} = - K_v V (\hat{\mathbf{m}} \cdot \hat{\mathbf{n}})^2$, which occurs when $\hat{\mathbf{m}} \cdot \hat{\mathbf{n}} = \pm 1$ and $\mathcal E_{K_v} = - K_v V$. The initial energy of the magnetic gyro is
\begin{equation}   \label{eq:energy-initial}
\mathcal E_{\rm tot} = - M_s V B \cos \theta_0 - K_v V.
\end{equation}
Here $\theta_0$ is the initial angle between $\hat{\mathbf m}=-\hat{\mathbf s}$ and $\mathbf B$; the initial rotational energy is neglected.  The energy of the state with the largest kinetic energy is
\begin{equation}   \label{eq:energy-max}
\mathcal E_{\rm tot} = \frac{1}{2} \mathbf{L}^T \cdot \boldsymbol{\mathcal{I}}^{-1} \cdot \mathbf{L} - M_s V B - K_v V.
\end{equation}
Setting these energies equal to one another gives an upper bound, because Gilbert damping dissipates energy during the transient (for the initial conditions used, $(\hat{\mathbf m}\!\cdot\!\hat{\mathbf n})^2 \approx 0.92$, so Eq.~(\ref{eq:energy-initial}) is accurate to $\approx 0.08 K_v V \approx 4 \times 10^{-18}$~J, small compared with $M_sVB(1-\cos\theta_0) \approx 8 \times 10^{-17}$~J); we find
\begin{align}   \label{eq:kinetic-energy-max}
\frac{1}{2} \mathbf{L}^T \cdot \boldsymbol{\mathcal{I}}^{-1} \cdot \mathbf{L} & \le M_s V B(1 - \cos \theta_0)  \nonumber \\
& = 2 M_s V B \sin^2 (\theta_0/2) .
\end{align}

The libration period quoted above follows from the magnetic torque on the magnetic gyro (when the spin is locked to the easy axis, and $S\ll L$ during the libration), $I_1\ddot\theta=-\mu B\sin\theta$, so that for small amplitude $\theta_0$ of the angle $\theta$ (i.e., for $\theta_0 \ll 1$~rad), $\omega_{\rm lib} = (\mu B/I_1)^{1/2} \approx 3.2 \times 10^{8}$~s$^{-1}$, and the libration period is $2\pi/\omega_{\rm lib} \approx 19.6$~ns for $B=0.5$~T.  If $\theta_0$ is not small, the libration period is
\begin{align*}
T_{\rm lib} &= 2 \sqrt{\frac{2 I_1}{\mu B}} \int\limits_{0}^{\theta_0} \frac{d\theta}{\sqrt{\cos\theta - \cos\theta_0}}\\
&\geq 2 \pi \sqrt{\frac{I_1}{\mu B}} = \frac{2\pi}{\omega_{\rm lib}} .
\end{align*}
[This follows from energy conservation, $\tfrac12I_1\dot\theta^2=\mu B(\cos\theta-\cos\theta_0)$, with $T_{\rm lib}=4\int_0^{\theta_0}d\theta/\dot\theta$.]  The ratio of the Larmor frequency $\omega_L=|\gamma|B$ to the libration frequency is $\omega_L/\omega_{\rm lib}=(|\gamma|BI_1/S)^{1/2} \approx 3.0 \times10^{2}$ at $0.5$~T and $0.95$ at $5~\mu$T; this is the dimensionless parameter that separates the pendulum-like regime from the gyroscopic regime.

\begin{figure}
\includegraphics[width=0.5\textwidth]{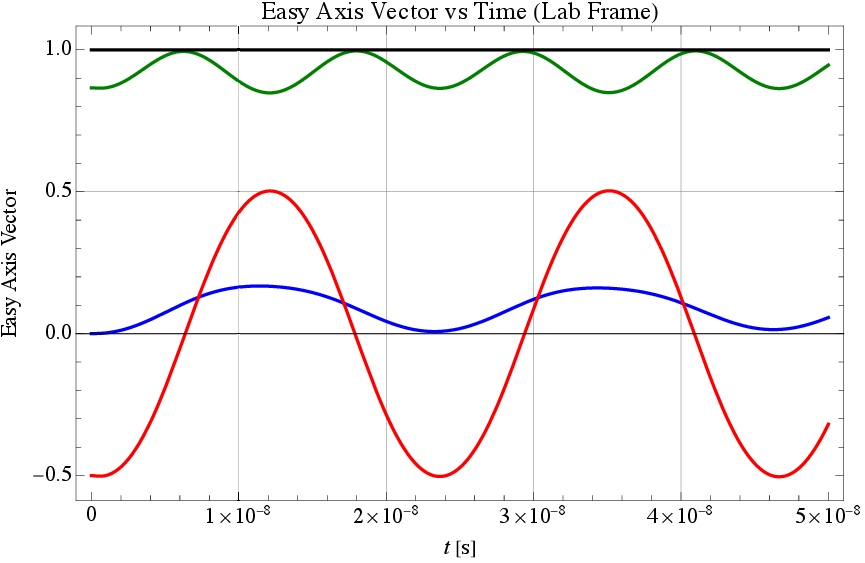}
\caption{\footnotesize  The easy axis ${\hat {\mathbf n}}(t)$ versus time in a magnetic field.   The blue curve is ${\hat n}_x(t)$, the red curve is ${\hat n}_y(t)$, and the green curve is ${\hat n}_z(t)$.  The black curve is the magnitude of ${\hat {\mathbf n}}(t)$, $n_x^2(t) + n_y^2(t) + n_z^2(t) = 1$.  The initial conditions and the parameters used in the calculation are the same as in Fig.~\ref{Fig_magnetic_needle_spin}.}
\label{Fig_magnetic_needle_easy_axis}
\end{figure}

\begin{figure}
\includegraphics[width=0.5\textwidth]{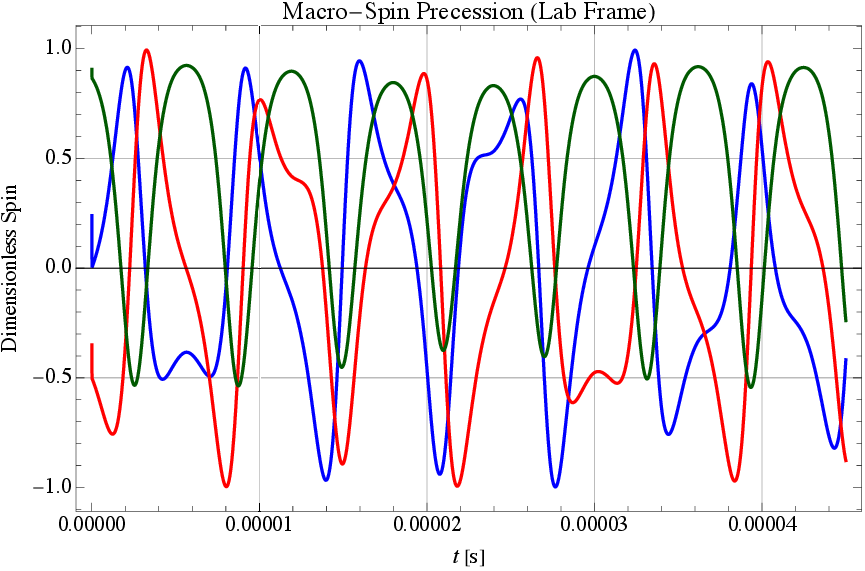}
\caption{Normalized spin, ${\mathbf s}(t)$, versus time with $B = 5 \, \mu$T, $t_f = 45 \, \mu$s, and the other parameters are same as used in Fig.~\ref{Fig_magnetic_needle_spin}.
The blue curve is $s_x(t)$, the red curve is  $s_y(t)$, and the green curve is $s_z(t)$.}
\label{Fig_magnetic_needle_spin_3}
\end{figure}

\begin{figure}
\includegraphics[width=0.5\textwidth]{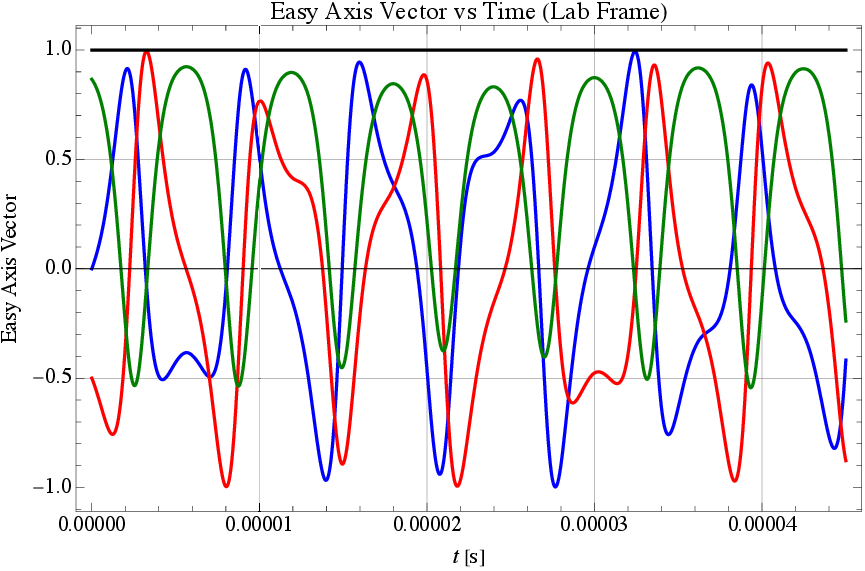}
\caption{The easy axis ${\hat {\mathbf n}}(t)$ versus time  with $B = 5 \, \mu$T, $t_f = 45 \, \mu$s, and the other parameters are same as used in Fig.~\ref{Fig_magnetic_needle_spin}.}
\label{Fig_magnetic_needle_easy_axis_3}
\end{figure}

Figure~\ref{Fig_magnetic_needle_easy_axis} plots the easy axis ${\hat {\mathbf n}}(t)$ versus time.  For times $t > \tau_G$ [see Eq.~(\ref{eq:LLG.tau})], ${\hat {\mathbf n}}(t)$ resembles the spin ${\hat {\mathbf s}}(t)$ in Fig.~\ref{Fig_magnetic_needle_spin}. However,  the direction of the easy axis, ${\hat {\mathbf n}}(t)$, is inverted relative to ${\hat {\mathbf s}}(t)$.  For $t > \tau_G$, the oscillation periods of ${\hat n}_x(t)$ and ${\hat n}_y(t)$ are roughly twice as large as the period of ${\hat n}_z(t)$. Note that the spin vector ${\mathbf s}(t)$ changes direction several times during the LLG damping period $\tau_G$. Each time the spin changes direction, the effective magnetic field in Eq.~(\ref{eq:B_eff}) also changes direction. This is because the anisotropy term, $\frac{2 K_v}{M_s} (\hat{\mathbf{m}} \cdot \hat{\mathbf{n}}) \hat{\mathbf{n}}$, is larger than the external magnetic field $\mathbf{B}$. Once the LLG damping process ends, the spin vector precesses around the effective magnetic field without the direction-change phenomenon occurring. If the spin changes direction an odd number of times, the direction of the easy axis is inverted relative to ${\hat {\mathbf s}}(t)$, as shown in Figs.~\ref{Fig_magnetic_needle_spin} and \ref{Fig_magnetic_needle_easy_axis}. Figures~\ref{Fig_magnetic_needle_spin_3} and \ref{Fig_magnetic_needle_easy_axis_3}, show what happens when the spin changes direction an even number of times, and the easy axis is aligned with the spin vector.

\begin{figure}
\includegraphics[width=0.5\textwidth]{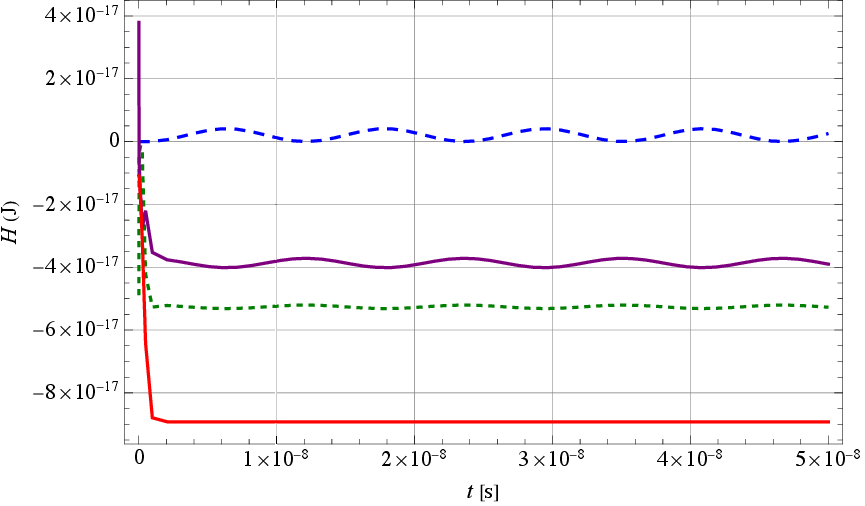}
\caption{The rotational kinetic energy (blue dashed), the anisotropy energy (green dashed), the Zeeman energy (purple dashed) and the total energy (red solid) are plotted versus time.  The initial conditions are the same as in Fig.~\ref{Fig_magnetic_needle_spin}.}
\label{Fig_magnetic_needle_energies}
\end{figure}

Figure \ref{Fig_magnetic_needle_energies} plots the rotational kinetic energy, the anisotropy energy, the Zeeman energy and the total energy (the sum of the previous three energies) versus time.  For times small compared to the LLG damping time $\tau_G$ there are significant changes in these energies, but for $t > \tau_G$ there are oscillations in the rotational kinetic energy, the anisotropy energy, the Zeeman energy, but the total energy remains constant.  The rotational kinetic energy and the Zeeman energy have oscillations that are 180 degrees out of phase.  The anisotropy energy has extremely small changes for $t > \tau_G$.

Figure~\ref{Fig.needle_spin} is a parametric plot of ${\mathbf s}(t)$.  For times $t < \tau_G$ the spin direction changes dramatically and then spirals until it begins to precess and nutate, as can be seen in Fig.~\ref{Fig_magnetic_needle_spin}.

\begin{figure}
\includegraphics[width=0.5\textwidth]{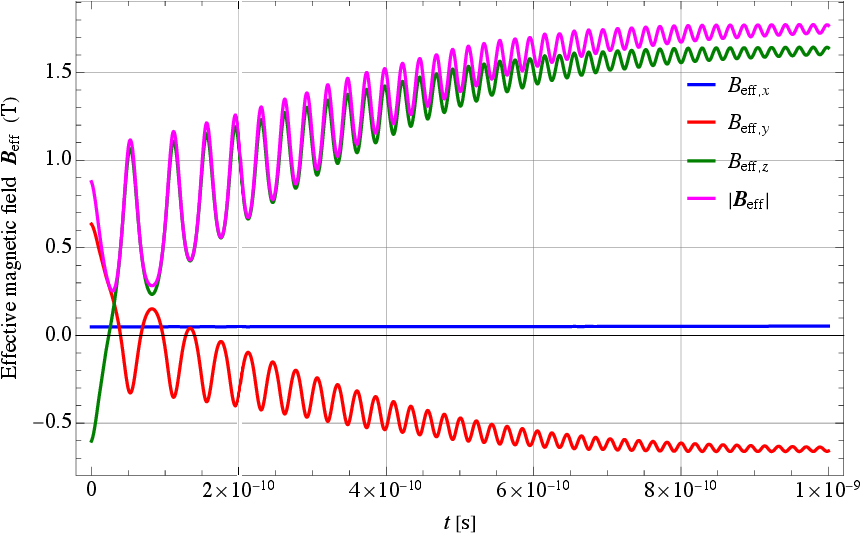}
\caption{The effective magnetic field $\mathbf{B}_{\text{eff}}(t)$ in Eq.~(\ref{eq:B_eff}) versus $t$ for $t < 1$~ns.  The external magnetic field and the initial conditions are the same as in Fig.~\ref{Fig_magnetic_needle_spin}.}
\label{Fig_B_eff_vs_t}
\end{figure}

Figure~\ref{Fig_B_eff_vs_t} plots the effective magnetic field $\mathbf{B}_{\text{eff}}(t)$ versus time for an external magnetic field of strength $B = 0.5$~T, and direction ${\hat {\mathbf B}} = (0.0995, 0., 0.995)$ in the time interval $0 < t < 1$~ns. The magnetic field component $B_{\text{eff},x}(t) \approx 0.05$~T is almost constant, while $B_{\text{eff},y}(t)$ and $B_{\text{eff},z}(t)$ vary rapidly with $t$.
Initially, $B_{\text{eff}, y}(0) = 0.64$~T and $B_{\text{eff}, z}(0) = - 0.6$~T; initially $B_{\text{eff}, y}(t)$ decreases, then vanishes at $t \approx 0.4$~ns and continues to oscillate as it becomes more negative, while $B_{\text{eff}, z}(t)$ initially increases, vanishes at $t \approx 0.02$~ns and continues to oscillate as it becomes more positive.  The amplitude of the oscillations of both $B_{\text{eff}, y}(t)$ and $B_{\text{eff}, z}(t)$ decrease with increasing time.  $B_{\text{eff}, y}(t)$ exhibits an overall downward trend, while $B_{\text{eff}, z}(t)$ exhibits an overall upward trend.  The magnitude of the effective magnetic field, $|\mathbf{B}_{\text{eff}}(t)|$, oscillates rapidly with time and increases with time, reaching a value of 1.75 T.  The average $|\mathbf{B}_{\text{eff}}(t)|$ over the time period $[0, \tau_G]$ where $\tau_G = |\alpha_G \, \gamma \, \mathbf{B}_{\text{eff}}(0)|^{-1} = 0.6$~ns [see Eq.~(\ref{eq:LLG.tau})] is roughly $\mathbf{B}_{\text{eff}}(0)$, and the LLG damping time is roughly 0.6~ns.

\begin{figure}
\includegraphics[width=0.45\textwidth]{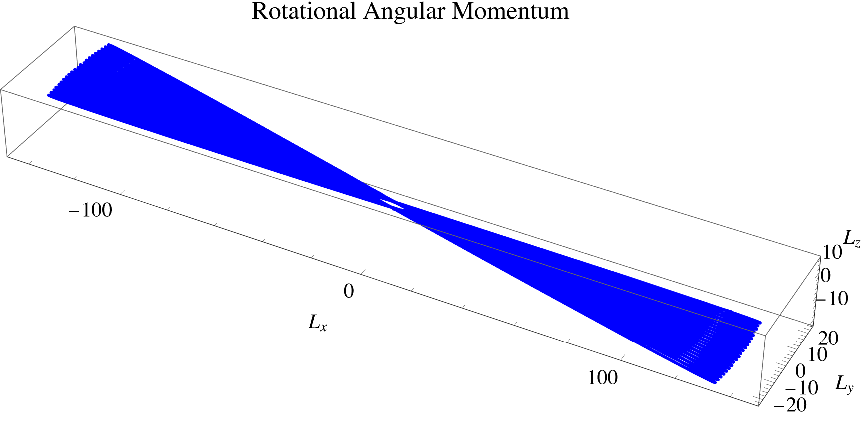}
\caption{\footnotesize  Parametric plot of the normalized rotational angular momentum, ${\mathbf l}(t)$. The integration time used is $t_f = 0.5 \, \mu$s.  The initial conditions and the parameters used in the calculation are the same as in Fig.~\ref{Fig_magnetic_needle_spin}.}
\label{Fig.needle_rotational_AM}
\end{figure}

Figure~\ref{Fig.needle_rotational_AM} is a parametric plot of the normalized rotational angular momentum, ${\mathbf l}(t)$.  The amplitude of the oscillations in $l_x(t)$ is larger than for $l_y(t)$ and $l_z(t)$.  Note that the rotational angular momentum ${\mathbf l}(t)$ is kept away from ${\mathbf l}(t) = {\bf 0}$.

\begin{figure} [!ht]
\includegraphics[width=0.45\textwidth]{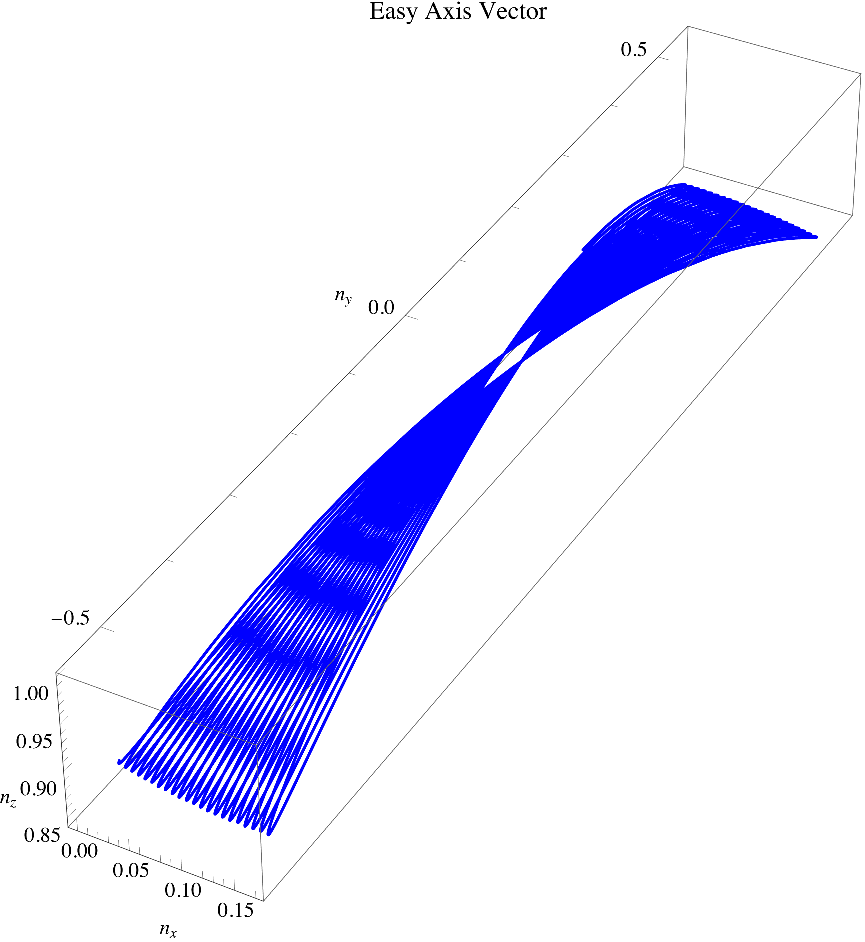}
\caption{\footnotesize  Parametric plot of the easy axis ${\hat {\mathbf n}}(t)$. The integration time used is $t_f = 0.5 \, \mu$s.  The black curve is $n_x^2(t) + n_y^2(t) + n_z^2(t) = 1$.  The initial conditions are the same as in Fig.~\ref{Fig_magnetic_needle_spin}.}
\label{Fig.needle_easy_axis}
\end{figure}

Figure~\ref{Fig.needle_easy_axis} is a parametric plot of the easy axis ${\hat {\mathbf n}}(t)$. The amplitude of oscillations of ${\hat n}_y(t)$ is large, and ${\mathbf n}(t)$ is kept away from $\hat{\bf z}$.

\begin{figure}
\includegraphics[width=0.5\textwidth]{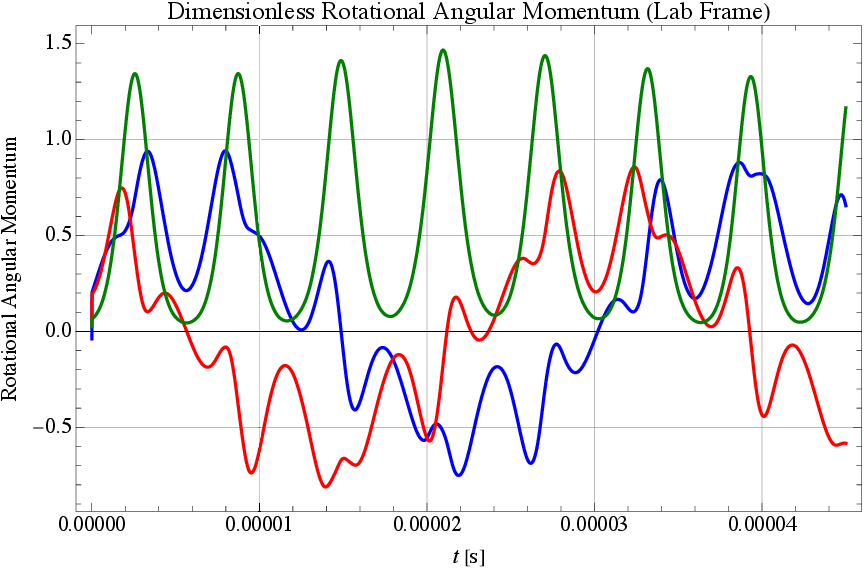}
\caption{Normalized rotational angular momentum, ${\mathbf l}(t)$, versus time in a magnetic field, with $B = 5 \, \mu$T, $t_f = 45 \, \mu$s, and the other parameters are same as used in Fig.~\ref{Fig_magnetic_needle_spin}.  The blue curve is $l_x(t)$, the red curve is  $l_y(t)$, and the green curve is  $l_z(t)$.}
\label{Fig_magnetic_needle_rotational_ang_momentum_3}
\end{figure}

Figures \ref{Fig_magnetic_needle_spin_3}, \ref{Fig_magnetic_needle_rotational_ang_momentum_3} and 
\ref{Fig_magnetic_needle_easy_axis_3} show the dynamics of ${\mathbf s}(t)$, ${\mathbf l}(t)$, ${\hat {\mathbf n}}(t)$ for $B = 5 \, \mu$T ($\ll 0.5$ T used above).  For this magnetic field strength, the Larmor precession rate, $\boldsymbol{\mu} \cdot \mathbf{B}/S$, is below the libration rate $(\boldsymbol{\mu} \cdot \mathbf{B}/I_1)^{1/2}$. The integration time used now is $t_f = 45 \, \mu$s  rather than the $t_f = 50$ ns used for Figs.~\ref{Fig_magnetic_needle_spin}-\ref{Fig_magnetic_needle_easy_axis}.

The spin dynamics ${\mathbf s}(t)$ are shown in Fig.~\ref{Fig_magnetic_needle_spin_3}.  Clearly, the oscillations are not harmonic.  The LLG dynamics is barely seen in the figure because $\tau_G = 0.4$ ns for $B = 5 \, \mu$T is much smaller than the integration time.  The amplitude of $s_x(t)$ is a bit smaller than those of $s_y(t)$ and $s_z(t)$. The period of oscillations after the first several oscillations is roughly 6 $\mu$s. 

The rotational angular momentum ${\mathbf l}(t)$ dynamics is shown in Fig.~\ref{Fig_magnetic_needle_rotational_ang_momentum_3}.  The component $l_x(t)$ is larger amplitude than $l_y(t)$ and  $l_z(t)$ during the first several oscillations, and the periods of the three components are roughly 6 $\mu$s after the first several oscillations.

Figure~\ref{Fig_magnetic_needle_easy_axis_3} plots the easy axis ${\hat {\mathbf n}}(t)$ versus time.  Now ${\hat {\mathbf n}}(t)$ and ${\hat {\mathbf s}}(t)$ are almost identical for times after the LLG dissipation dynamics ends.

\begin{figure}
\includegraphics[width=0.5\textwidth]{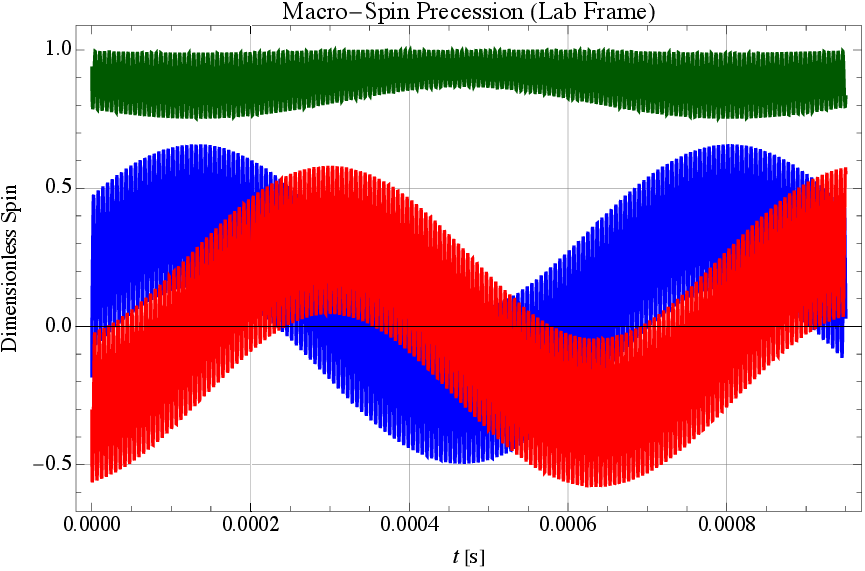}
\caption{Normalized spin, ${\mathbf s}(t)$, versus time with $B = 50$ nT, $t_f = 1$ ms, and the other parameters are same as used in Fig.~\ref{Fig_magnetic_needle_spin}.
The blue curve is $s_x(t)$, the red curve is  $s_y(t)$, and the green curve is $s_z(t)$.}
\label{Fig_magnetic_needle_spin_4}
\end{figure}

Figure \ref{Fig_magnetic_needle_spin_4} shows the dynamics of ${\mathbf s}(t)$ for $B = 50$ nT and an integration time of $t_f = 1$ ms. There are two oscillation periods: one at approximately $650 \, \mu$s, which is inversely proportional to the square root of the magnetic field, and one at about $6 \, \mu$s.  The rotational angular momentum motion (not shown) has the same two periods.

\begin{figure}
\includegraphics[width=0.5\textwidth]{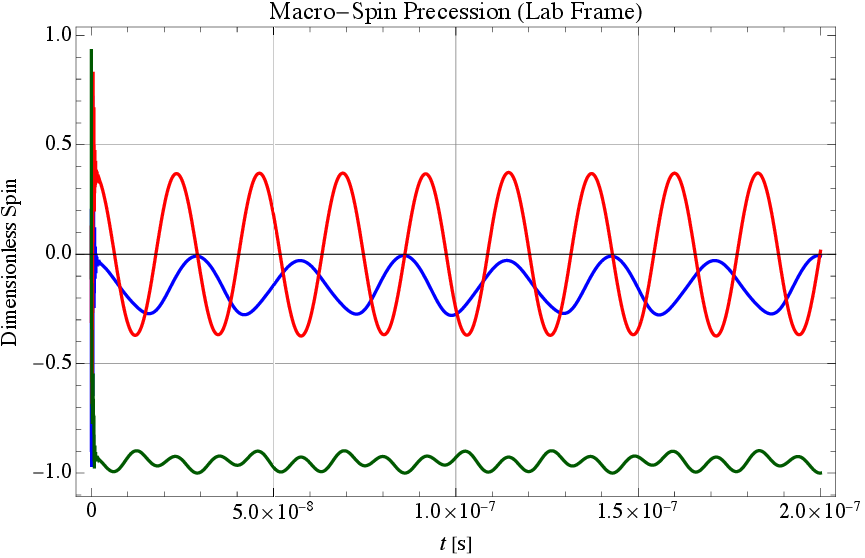}
\caption{\footnotesize  Normalized spin, ${\mathbf s}(t)$, versus time with $E = 5.0 \times  10^{7}$ V/m, direction ${\hat {\mathbf E}} = \hat{\mathbf{x}}$, and the same magnetic field as in Fig.~\ref{Fig_magnetic_needle_spin}.  The blue curve is $s_x(t)$, the red curve is  $s_y(t)$, and the green curve is  $s_z(t)$.  The initial conditions are the same as in Fig.~\ref{Fig_magnetic_needle_spin}.}
\label{EB.needle_spin}
\end{figure}

\begin{figure}
\includegraphics[width=0.5\textwidth]{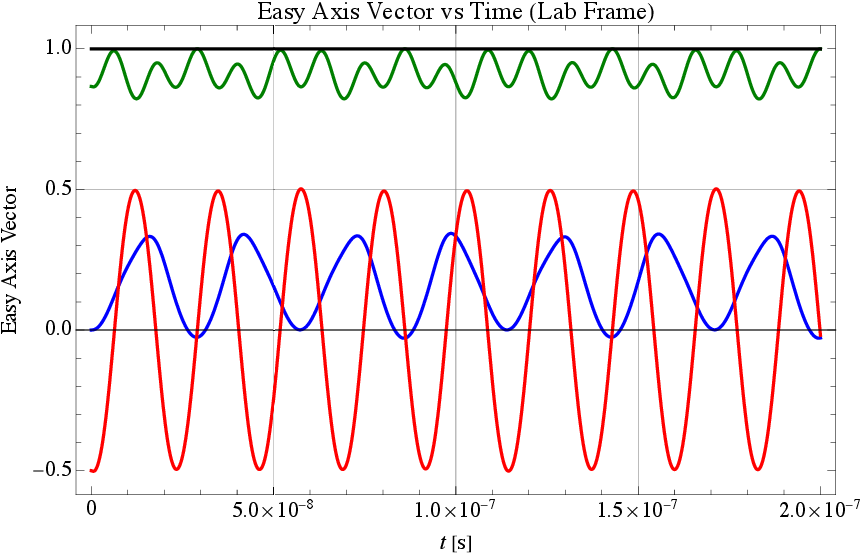}
\caption{\footnotesize  Easy axis ${\hat {\mathbf n}}(t)$ versus time with the same magnetic and electric fields as in Fig.~\ref{EB.needle_spin}.  The blue curve is $n_x(t)$, the red curve is  $n_y(t)$, and the green curve is  $n_z(t)$.  The initial conditions are the same as in Fig.~\ref{Fig_magnetic_needle_spin}.}
\label{EB.needle_easy_axis}
\end{figure}

Figure~\ref{EB.needle_spin} shows the dynamics of the spin ${\mathbf s}(t)$ in the same strong magnetic field as used in Fig.~\ref{Fig_magnetic_needle_spin} and in a strong static electric field of strength $E = 5.0 \times  10^{7}$ V/m, direction ${\hat {\mathbf E}} = \hat{\mathbf{x}}$ (nearly perpendicular to $\mathbf B$, since $\hat{\mathbf B}\!\cdot\!\hat{\mathbf x}=0.0995$). The electrostatic energy scale $p_E E^2 \approx 1.4 \times 10^{-17}$~J is smaller than the Zeeman scale $\mu B\approx4.0\times10^{-17}$~J at 0.5~T by a factor of about 3, i.e., $E \approx 0.6 E_c$, where $E_c \approx 8.5 \times 10^{7}$~V/m is the crossover field of Eq.~(\ref{eq:Ec}), so that the polarization torque is comparable to, but smaller than the magnetic torque. The magnetoelectric field, $\alpha E \approx 10^{-6}$~T, and the motional field, $\sim Ev/c^2$, are negligible in comparison. The polarization torque tends to align the long axis with $\mathbf E$ [Eq.~(\ref{eq:pE}) and $U_E$]; since the needle starts at $\hat{\mathbf n}\!\cdot\!\hat{\mathbf E}=0$, an unstable maximum of $U_E$, it is driven away from this orientation toward $\mathbf E$, but the magnetic torque keeps the axis near $\mathbf B_{\rm eff}$: $n_x(t)$ (Fig.~\ref{EB.needle_easy_axis}) oscillates between $\approx0$ and $\approx0.34$, i.e., the axis is tilted toward $\mathbf E$ by up to $\approx20^\circ$, on the $\approx20$--$30$~ns time scale set by the magnetic and electrostatic libration periods [$2\pi(I_1/\mu B)^{1/2}\approx20$~ns and $2\pi(I_1/p_EE^2)^{1/2}\approx33$~ns].

\begin{figure}
\includegraphics[width=0.5\textwidth]{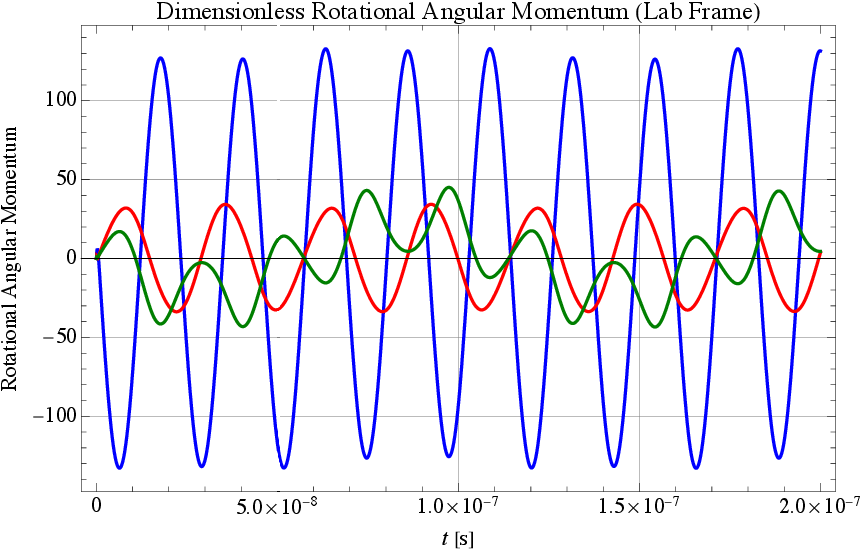}
\caption{\footnotesize  Normalized rotational angular momentum ${\mathbf l}(t)$ versus time with the same magnetic and electric fields as in Fig.~\ref{EB.needle_spin}.  The blue curve is $l_x(t)$, the red curve is  $l_y(t)$, and the green curve is  $l_z(t)$.  The initial conditions are the same as in Fig.~\ref{Fig_magnetic_needle_spin}.}
\label{EB.needle_rotational_AM}
\end{figure}

Figure~\ref{EB.needle_rotational_AM} plots the rotational angular momentum versus time. As in Fig.~\ref{Fig_magnetic_needle_rotational_ang_momentum}, $l_x$ dominates (peak $\approx130$), but $l_y(t)$ and $l_z(t)$ are about twice as large as without the electric field (up to $\approx40$).

Figure~\ref{EB.needle_easy_axis} plots the easy magnetization axis. Compared with Fig.~\ref{Fig_magnetic_needle_easy_axis}, $\hat n_x$ is now also systematically positive (the axis leans toward $\mathbf E$), while $\hat n_z$ remains $\gtrsim 0.8$.  In contrast to Fig.~\ref{Fig_magnetic_needle_easy_axis}, $\hat n_z$ has more than one period, and the periods appear to be non-commensurate.  The temporal duration between peaks of $\hat n_z$ varies from peak to peak, but $\hat n_y$ appears to be periodic (note that the peaks are broader then the minima).

\begin{figure}
\includegraphics[width=0.5\textwidth]{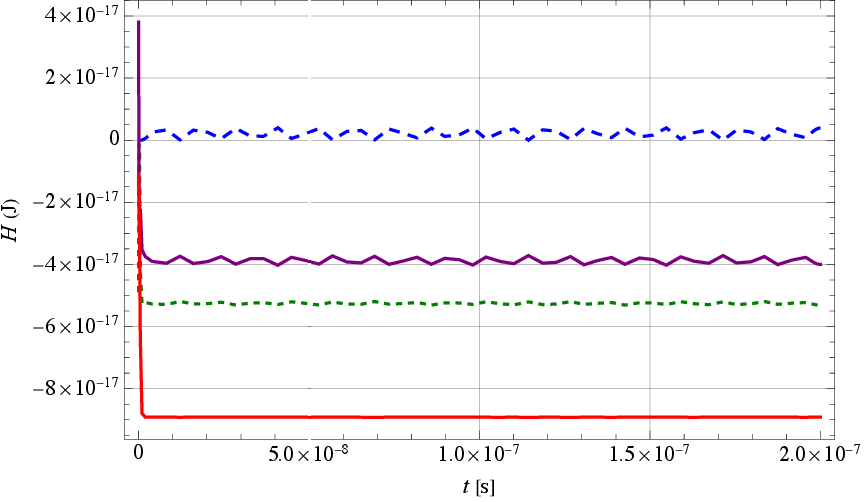}
\caption{The rotational kinetic energy (blue dashed), the anisotropy energy (green dashed), the electric-magnetic energy (purple) and the total energy (red) are plotted versus time with the same magnetic and electric fields as in Fig.~\ref{EB.needle_spin}.  The initial conditions are the same as in Fig.~\ref{Fig_magnetic_needle_spin}.}
\label{Fig_magnetic_needle_energies.EB}
\end{figure}

Figure \ref{Fig_magnetic_needle_energies.EB} plots the rotational kinetic energy, the anisotropy energy, the electric-magnetic energy $H_{\rm EM}$ [which contains the induced-dipole energy $-\frac12p_E(\mathbf E\!\cdot\!\hat{\mathbf n})^2$ and the Zeeman energy] and the total energy. The figure is similar to Fig.~\ref{Fig_magnetic_needle_energies}, except for the time scale, which is about four times larger.

\begin{figure}
\includegraphics[width=0.5\textwidth]{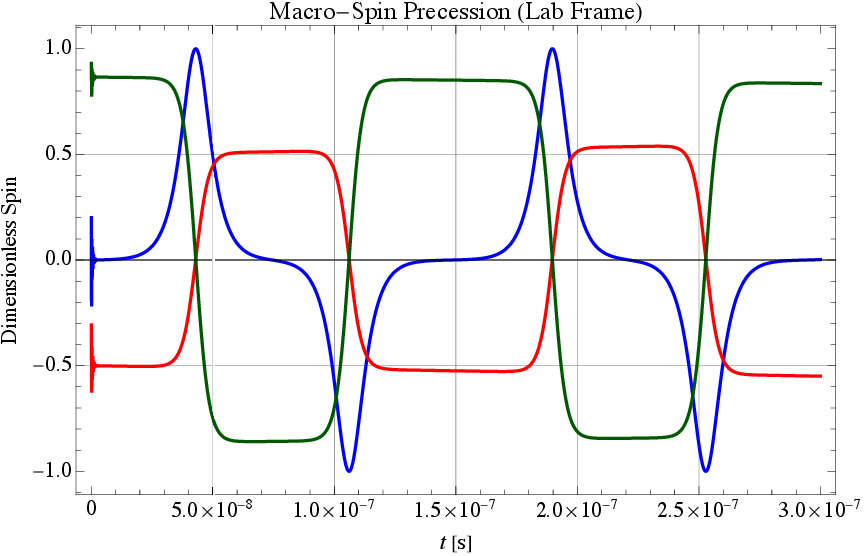}
\caption{\footnotesize  Normalized spin, ${\mathbf s}(t)$, versus time with $E = 5.0 \times  10^{7}$ V/m, direction ${\hat {\mathbf E}} = \hat{\mathbf{x}}$, and the same magnetic field as in Fig.~\ref{Fig_magnetic_needle_spin_3} (i.e., with $B = 5 \, \mu$T).  The blue curve is $s_x(t)$, the red curve is  $s_y(t)$, and the green curve is  $s_z(t)$.  The initial conditions are the same as in Fig.~\ref{Fig_magnetic_needle_spin}.}
\label{EB_low.needle_spin}
\end{figure}

\begin{figure}
\includegraphics[width=0.5\textwidth]{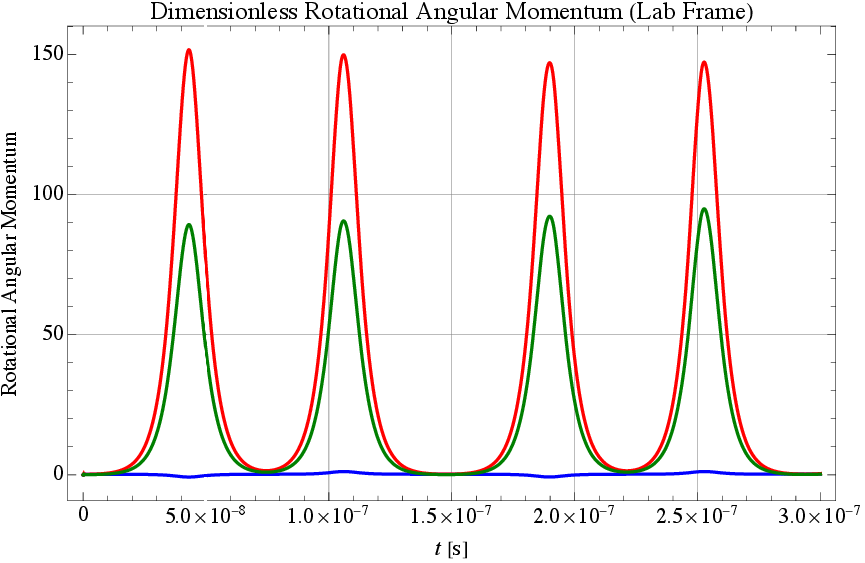}
\caption{\footnotesize  Normalized rotational angular momentum ${\mathbf l}(t)$ versus time with the same magnetic and electric fields as in Fig.~\ref{EB_low.needle_spin}.  The blue curve is $l_x(t)$, the red curve is  $l_y(t)$, and the green curve is  $l_z(t)$.  The initial conditions are the same as in Fig.~\ref{Fig_magnetic_needle_spin}.}
\label{EB_low.needle_rotational_AM}
\end{figure}

\begin{figure}
\includegraphics[width=0.5\textwidth]{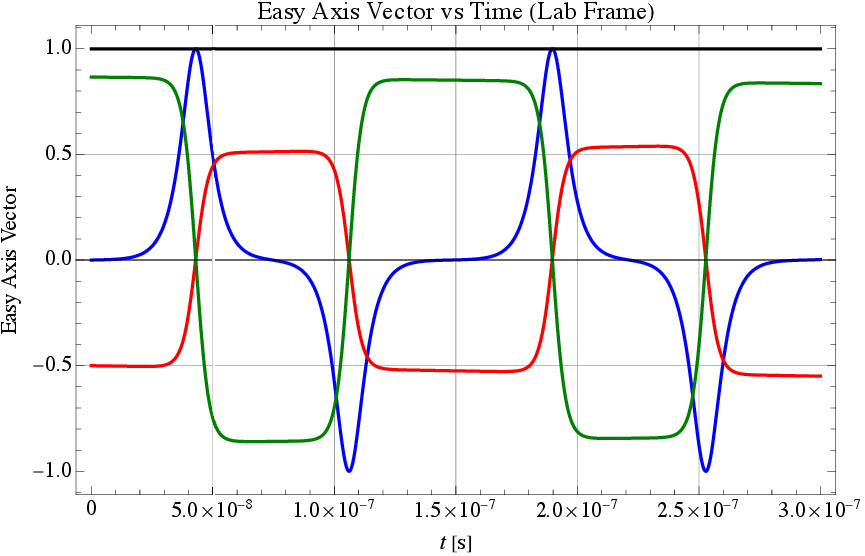}
\caption{\footnotesize  Easy axis ${\hat {\mathbf n}}(t)$ versus time with the same magnetic and electric fields as in Fig.~\ref{EB_low.needle_spin}.  The blue curve is $n_x(t)$, the red curve is  $n_y(t)$, and the green curve is  $n_z(t)$.  The initial conditions are the same as in Fig.~\ref{Fig_magnetic_needle_spin}.}
\label{EB_low.needle_easy_axis}
\end{figure}

\begin{figure}
\includegraphics[width=0.45\textwidth]{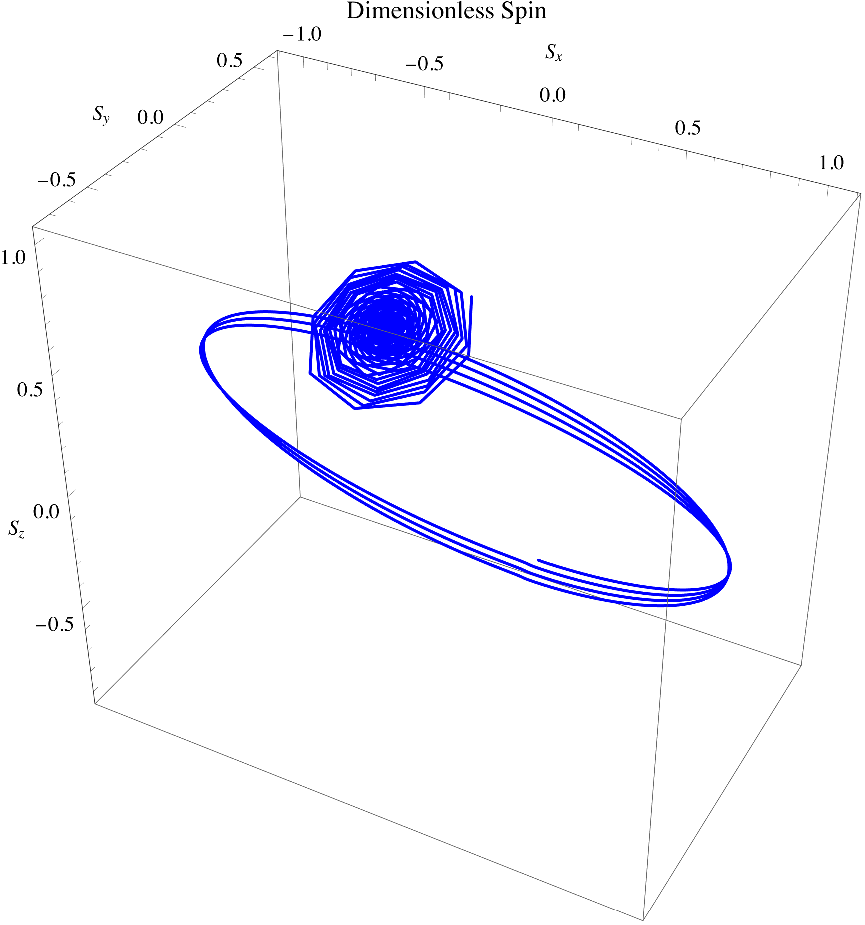}
\caption{\footnotesize  Parametric plot of the normalized spin, ${\mathbf s}(t)$ with the same magnetic and electric fields as in Fig.~\ref{EB_low.needle_spin}.  The integration time used is $t_f = 0.5 \, \mu$s.  The initial conditions and the parameters used in the calculation are the same as in Fig.~\ref{Fig_magnetic_needle_spin}.}
\label{Fig.needle_spin.B_low&E}
\end{figure}

Figures~\ref{EB_low.needle_spin}--\ref{Fig.needle_spin.B_low&E} show the same electric field [$E = 5.0 \times  10^{7}$ V/m, direction ${\hat {\mathbf E}} = \hat{\mathbf{x}}$] but with a small magnetic field strength, $B=5~\mu$T, so the magnetic torque is smaller than the electric torque. Figure \ref{EB_low.needle_spin} shows the spin dynamics.  We see that, after the LLG damping time $\tau_G$, the spin behaves very differently from the other cases plotted above.  The shape of the $s_x(t)$ is spiked, and $s_y(t)$ and $s_z(t)$ are flat on top and bottom. Figure \ref{EB_low.needle_rotational_AM} shows the rotational angular momentum.  Since $\mathbf N_{\rm elec}\perp\mathbf E$, the component of the total angular momentum along $\mathbf E$ is conserved when the magnetic torque is negligible, and since $|{\bf s}(t)| = 1$ and the initial rotational angular momentum is small,  $|l_x(t)| < 1$, while $l_y(t)$ and $l_z(t)$ are large in comparison. Figure \ref{EB_low.needle_easy_axis} shows the easy-axis dynamics.  For $t > \tau_G$, ${\hat {\mathbf n}}(t)$ looks just like ${\hat {\mathbf s}}(t)$.

\section{Summary and Conclusion}  \label{sec:Summary}

We have formulated the coupled spin-rotation dynamics of a single-domain ferromagnetic needle in combined static magnetic and electric fields, using quaternion kinematics, in the laboratory and body-fixed frames, and in dimensionless form. The new results presented in this article are as follows:
\begin{enumerate}
\item The electric-field couplings of a magnetic gyroscope: the shape-dependent polarization torque of the conducting needle, Eq.~(\ref{eq:polarization_torque}), the linear magnetoelectric coupling, Eq.~(\ref{eq:magnetoelectric}), and the motional spin-orbit coupling $-\frac{\gamma}{c^2} \mathbf{S} \cdot [\mathbf{E} \times (\mathbf{v}_{\text{cm}} + \boldsymbol{\Omega} \times \mathbf{r}_{\text{eff}})]$, in which a displacement $\mathbf r_{\rm eff}$ of the spin density from the center of rotation makes the motional field depend on the rotation of the needle (in our examples the motional spin-orbit coupling vanishes).

\item The Gilbert dissipation function depends on the rate of change of the magnetization relative to the lattice. The LLG (i.e., Gilbert) torque and the internal anisotropy torque therefore act with equal and opposite signs on the spin and on the lattice, and therefore do not affect the total angular momentum $\mathbf J = \mathbf S+\mathbf L$. In the body frame the rotation enters the spin dynamics as the Barnett field $\boldsymbol\Omega/\gamma$.

\item For a 120-nm cobalt needle in 0.5~T the anisotropy field ($\approx1.3$~T) exceeds the applied field, and the spin reverses several times during the effective Gilbert decay transient time ($\tau_G \approx 0.6$~ns). The parity of the number of reversals determines whether the easy axis ends parallel or antiparallel to the spin. The energy released during the transient is converted partly into rotation of the needle, with $\frac12\mathbf L\cdot\boldsymbol{\mathcal I}^{-1}\cdot\mathbf L\le 2M_sVB\sin^2(\theta_0/2)$.

After the transient the spin is locked to the easy axis and the needle librates about the field (precession with nutation at twice the frequency), with a period of about 20~ns. This period agrees with $2\pi(I_1/\mu B)^{1/2}$ and is much longer than the Larmor period. The total energy is conserved after the transient, and the Zeeman and rotational kinetic energies oscillate out of phase.

\item For a magnetic field of $5 \, \mu$T the magnetic gyro has a period of about $6 \, \mu$s. However, when the magnetic field decreases to $50$ nT the gyro's period increases to approximately $650 \, \mu$s.  This oscillation period is inversely proportional to the square root of the magnetic field.  Additionally, there is an oscillation with a shorter period of about $6 \, \mu$s.

\item The polarization torque tends to align the needle axis with $\mathbf E$. For $E=5\times10^{7}$~V/m nearly perpendicular to $\mathbf B$, $E\approx0.6E_c$ at 0.5~T and the electric field tilts the needle toward $\mathbf E$ by up to $\approx20^\circ$; at $B=5~\mu$T, $E\approx190E_c$, the polarization torque dominates, the needle swings into the field direction, and rotational kinetic and electrostatic energy are exchanged at constant total energy. The magnetoelectric and motional fields are of order $1\ \mu$T or smaller (for a cobalt ellipsoid it is zero by symmetry). The induced-dipole description is expected to hold for $E_0 \lesssim 2\times 10^{8}$~V/m, because the local field at the tips is enhanced by $1/N_\parallel \approx 13$ (Sec.~\ref{subsec:screening}). A conducting needle can be used as a magnetometer only if $E \ll E_c=[M_sB/\varepsilon_0 (N_\parallel^{-1}-N_\perp^{-1})]^{1/2}$ ($\approx 8.5 \times 10^{7}$~V/m at 0.5~T), a criterion that is independent of the size of the needle (it scales as $B^{1/2}$ and depends on the aspect ratio). Stray electric fields must therefore be controlled when such a needle is used to measure static or low-frequency magnetic fields.
\item When measuring a static or low-frequency magnetic field with a magnetic gyroscope, one must ensure that electric fields are not present because an electric field can significantly affect the dynamics via the electrostatic polarization torque [see Eq.~(\ref{eq:polarization_torque})].

\item In the body-fixed frame the static electric field appears as a field $\mathbf E_b(t)=\mathbf R^T(t)\mathbf E$ rotating at the rotation frequency. For $\Omega/2\pi\sim1$~GHz the conduction electrons of the cobalt needle still screen it essentially perfectly ($|\varepsilon| \sim 3 \times 10^{8}$), so the induced dipole is the static one, $\mathbf p = \boldsymbol\chi \mathbf E_b$, while the magnetic field is not screened (skin depth $\gtrsim 0.4~\mu$m $\gg$ needle size). The field at the tips is enhanced by $1/N_\parallel\approx 13$, which limits the validity of the model to $E_0 \lesssim 2\times10^{8}$~V/m (field emission).
\item Collisions of the free electrons in the conducting needle provide an electric analog of Gilbert damping: in the body frame the induced dipole obeys a Debye relaxation equation with $\tau_{\chi,i} = \varepsilon_0 /(N_i\sigma)$, derived from the Joule dissipation function, $\mathcal R_E = |d\mathbf p/dt -\boldsymbol\Omega \times \mathbf p|^2/(2V\sigma)$, which has the same frame-covariant form as the Gilbert dissipation function. The resulting lag torque, together with eddy-current friction, damps the rotation on a time scale of $\sim10$~ms (eddy currents at 0.5~T) to $\sim3$~s (Joule friction at $E=5\times10^{7}$~V/m) for the cobalt needle, much longer than the Gilbert time $\tau_G \approx 0.6$~ns, and is negligible on the time scales of our simulations.
\end{enumerate}


\end{document}